
\magnification=1200
\hsize=6.4 truein \vsize=8.9 truein

\tolerance=10000
\def\folio{\ifnum\pageno=1\nopagenumbers\else\number\pageno\fi}
%
        
        \skip\footins=30pt
        \def\footnoterule{\kern-3pt
        \hrule width \the\hsize \kern 2.6pt}
\baselineskip 12pt minus 1pt
\parskip=\medskipamount
\def\nulltest{}    
\def\headline{}    
\def\headlineb{}   
\def\makeheadline{\ifnum\pageno=1 \else
  \baselineskip 12pt \ifx \headline \nulltest \else
     \line{\headline} \advance \vsize by -\baselineskip \advance
     \vsize by -14pt \fi
   \ifx \headlineb \nulltest \else
     \line{\headlineb} \advance \vsize by -\baselineskip \fi
   \ifx \headline \nulltest \else \vskip 14pt \fi
  \fi}   

\def\refpar{\parshape=2 0truein 6.5truein 0.3truein 6.2truein}
\def\ref#1;#2;#3;#4{{\par\refpar #1, {\it #2}, {\bf #3}, #4}}
\def\prep#1;#2{{\par\refpar #1, #2}}
\def\book#1;#2;#3{{\par\refpar #1, {\it #2} #3}}

\def\ltsima{$\; \buildrel < \over \sim \;$}
\def\lsim{\lower.5ex\hbox{\ltsima}}
\def\gtsima{$\; \buildrel > \over \sim \;$}
\def\gsim{\lower.5ex\hbox{\gtsima}}

\def\kms{\,{\rm km\,s^{-1}}}

\def\pp{\dimen0=\hsize
        \advance\dimen0 by -1truecm
        \par\parshape 2 0truecm \dimen0 1truecm \dimen0 \noindent}
\def\ltsima{$\; \buildrel < \over \sim \;$}
\def\lsim{\lower.5ex\hbox{\ltsima}}
\def\gtsima{$\; \buildrel > \over \sim \;$}
\def\gsim{\lower.5ex\hbox{\gtsima}}
\vbox to 0.25in{ }
\vskip .2truein
\centerline{\bf COLD DARK MATTER I:}
\centerline{\bf THE FORMATION OF DARK HALOS}
\vskip .2truein
\centerline{\bf James M. Gelb$^\dagger$ and Edmund Bertschinger}
\vskip .1truein
\centerline{Department of Physics,}
\centerline{Massachusetts Institute of Technology,}
\centerline{Cambridge, MA 02139}
\vskip .1truein
\centerline{$^\dagger$Present address:}
\centerline{NASA/Fermilab Astrophysics Center,}
\centerline{Fermi National Accelerator Laboratory,}
\centerline{P.O. Box 500, Batavia, IL 60510}
\vskip .2truein
\centerline{\bf ABSTRACT}
\vskip .2truein
{\baselineskip 12pt minus 1pt
{We use numerical simulations
of critically-closed cold dark matter (CDM) models
to study the effects of numerical resolution on observable
quantities.
We study simulations with up to $256^3$
particles
using the particle-mesh (PM) method and with up to
$144^3$ particles using
the adaptive particle-particle--particle-mesh (P$^3$M) method.
Comparisons of galaxy halo distributions are
made among the various simulations.  We also compare distributions
with observations and we explore methods
for identifying halos, including a new algorithm that finds all
particles within closed contours of the smoothed density field
surrounding a peak.
The simulated halos show more substructure than
predicted by the Press-Schechter theory.
We are able to rule out all $\Omega=1$ CDM models for
linear amplitude $\sigma_8\gsim 0.5$ because the simulations produce
too many massive halos compared with the observations.
The simulations also produce too many low mass halos.
The distribution of halos characterized
by their circular velocities for the P$^3$M simulations
is in reasonable agreement with the observations
for $150\kms\lsim V_{\rm circ}\lsim 350\kms$.}}
\vskip .3truein
\noindent{\it Subject headings:}
cosmology: theory --- dark matter --- galaxies: clustering ---
galaxies: formation
\vskip .3truein
\vskip .2truein
\vfill
\eject
\vskip .3truein
\centerline{\bf 1. INTRODUCTION}
This paper is part of a two part series testing the
cold dark matter (CDM) model of galaxy formation assuming
a critically-closed universe, $\Omega=1$.
These papers focus on the formation and clustering of halos
in cosmologically significant volumes of space (cubes
of length $\gsim$ 50 Mpc on a side) with sufficient
mass resolution and length resolution (force softening and box size)
to resolve thousands of individual halos.
The goal is not to study large scale structure
($\gsim 200$ Mpc; {\it e.g.} Park 1990).
Rather, the goal is to study spatial and velocity statistics
on scales $\sim 1-10$ Mpc
using candidate galaxy halos identified in the nonlinear,
evolved density field.
A principal goal of both papers is to determine if there exists a
linear normalization of the initial fluctuation power spectrum (a free
parameter in the theory) that satisfies observational constraints
on galaxy masses, clustering, and velocities, and galaxy cluster
multiplicity functions.

The principle goal of {\it this} paper is to understand the properties
of dark halos that form in cosmologically significant volumes of space
in the CDM model.  Specifically, we want to understand the
sensitivity of halo formation and halo properties to
numerical resolution.  We identify which properties of halo
formation
({\it e.g.} distributions of halo mass and circular velocity)
are particularly sensitive to such parameters as box size, force resolution,
mass resolution, and methods for identifying
halos.

Other workers
have studied the formation of dark halos in the CDM scenario
in volumes of space much greater than $(100~{\rm Mpc})^3$ by using
approximate methods for identifying galaxies as individual
particles ({\it e.g.} Davis et al. 1985).
Still others have studied volumes of space much smaller than $(100~{\rm
Mpc})^3$
with relatively high mass and force resolution
({\it e.g.} Frenk et al. 1988).  Small volumes of space
do not contain long wavelengths in the initial conditions
which {\it may} affect halo formation (studied in this paper) and
which {\it do} affect clustering (Gelb \& Bertschinger 1994, hereafter
Paper II).
The larger volumes
of space simulated with relatively high numerical resolution
presented in this paper
also yield better statistics since more halos form than in smaller volumes.

{}From our efforts, based on over one thousand IBM 3090 supercomputer-hours
applied to more than a dozen large simulations, we gain
insight into dynamic range by systematically isolating various effects.
We demonstrate which halo properties, if any,
converge with increasing resolution up to practical limits
using present-day supercomputers.  These dynamic range studies
are important for future workers who need to choose a particular
set of simulation parameters for a particular problem in galaxy
formation.

By comparing the distribution of halo masses
with estimates from observed galaxies,
we show that the simulations produce too many massive halos.
In Paper II,
focusing on the spatial
and velocity statistics of the halos, we consider the possibility
that the overly massive halos represent clusters of galaxies
(Katz \& White 1993; Evrard, Summers, \& Davis 1994).
Because our simulations do not include gas dynamical dissipation,
it is possible that the dark matter halos we identify have clustering
properties different from the luminous galaxies that would form if
we properly simulated all of the physics of galaxy formation.  To
minimize the uncertainty caused by our lack of dissipative physics,
we try to employ tests that should not depend strongly on the
relation between dark halos and luminous galaxies.  For the same reason,
in Paper II we explore several different prescriptions for
galaxy formation and we discuss cosmological
N-body simulations employing gas dynamics ({\it e.g.} Cen \& Ostriker 1992a,b;
Katz, Hernquist, \& Weinberg 1992).

The N-body simulations follow the nonlinear gravitational
clustering (in an expanding universe)
of particles representing collisionless clouds
of dark matter.  The simulations utilize
between $64^3$ (262144) and $256^3$ (16777216) particles in a
universe with $\Omega=1$ and $H_0=50~\kms~{\rm Mpc}^{-1}$.  All
distances are given in units Mpc rather than $h^{-1}\,$Mpc.
Most of the simulations are computed in cubes of length
51.2 Mpc on a side (box sizes are comoving).
As we show in Paper II, this volume is too small
to accurately measure galaxy clustering, although it allows
one to resolve thousands of individual halos with hundreds to thousands of
particles per typical Milky Way-sized halo.
(We do, however, compute a few simulations in boxes of order 100 Mpc
on a side in order to study galaxy clustering and small-scale peculiar
velocities in Paper II.)

Our simulations employ both the particle-mesh (PM) method (Hockney \&
Eastwood 1982) and the adaptive particle-particle--particle-mesh (P$^3$M)
method (Couchman 1991).  For a review of N-body methods in cosmology
see Bertschinger (1991).  Bertschinger \& Gelb (1991) provide an overview
of the numerical aspects of this work.  Gelb (1992) provides many technical
details and is the basis of these papers.

In the remainder of this introduction we discuss briefly three key issues
relevant for cosmological simulations of galaxy halo formation: force
resolution, halo identification, and the normalization of the power
spectrum.  In $\S$ 2 we use the cumulative mass fraction of halos to study
the effects of numerical resolution on halo formation, and we compare the
simulations with the Press-Schechter (1974) theory.  In $\S$ 3 we
explore circular velocity profiles and introduce observational data.
In $\S$ 4 we compare the number of halos, characterized by their circular
velocities, with observations.  Separate sub-sections
are included for high mass halos and for low mass halos.
Conclusions and a summary are given in $\S$ 5.

\vskip .2truein
\centerline{\it 1.1. Force Resolution}

An important ingredient in N-body simulations is force resolution.
We characterize the force softening in the simulations (with
particle mass $m_{\rm part}$) by the comoving pair separation $r=R_{1/2}$ such
that
$r^2F_r/(Gm_{\rm part}^2)=1/2$, {\it i.e.} where the radial component of the
force between
two particles is half its Newtonian value.  For the PM simulations
$R_{1/2}\approx 1.4$ grid cells (Gelb 1992 chapter 2).  For P$^3$M simulations
with a Plummer force law characterized by a softening $\epsilon$, {\it i.e.}
with
$F_r=Gm_{\rm part}^2r/(r^2+\epsilon^2)^{3/2}$, $R_{1/2}\approx 1.305\epsilon$.
The shape of the PM softening is slightly different from a Plummer law,
but in each case the appropriate force law (inverse square or Plummer)
is matched accurately (to better than 2\% rms) for $r>2R_{1/2}$.
There is additionally a small transverse component of the force due to
PM grid anisotropies.  Force errors are minimized using a suitable Green's
function; see Bertschinger (1991), Gelb (1992), and Hockney \& Eastwood
(1982).

For economy of notation and ease of reference
we refer to the simulations as follows: CDMn($N$,$L$,$R_{1/2}$).
Following Gelb (1992), we number the simulations from $n=1$ to 16.
The numbers in parentheses indicate the following simulation parameters:
1) $N$ particles, 2) a comoving box of length $L$
Mpc on a side, and 3) a comoving force softening length of $R_{1/2}$ kpc.
For example, CDM1(128$^3$,51.2,280) uses $128^3$ particles,
a $(51.2~{\rm Mpc})^3$ box, and a comoving
force softening length of 280 kpc.
The two P$^3$M simulations discussed in these papers
use $R_{1/2}=52$ kpc comoving ($\epsilon=40$ kpc) and
$R_{1/2}=85$ kpc comoving ($\epsilon=65$ kpc).
The other simulations are low force resolution PM simulations with
$R_{1/2}\ge 190$ kpc comoving.

We summarize the simulation parameters in Table~1.
The entries are the following:
simulation number, particle-mesh grid, particle mass,
starting expansion factor, number of timesteps
to $\sigma_8=1$,
energy conservation relative to change in gravitational potential
energy (see Gelb 1992 chapter 2), computer hours consumed,
initial conditions identifier.
Simulations with
the same initial conditions identifier use equivalent sets
of random numbers, {\it i.e.} they are generated
from the same set of random numbers scaled to
the appropriate power spectrum (see Gelb 1992 chapter 3).

We use a time-centered leapfrog scheme (Hockney \& Eastwood 1982)
to advance the particles.
All of the simulations are integrated using equal steps in expansion
factor $a$,
except CDM12, which uses equal
steps in $a^\alpha$ with $\alpha=0.5$, as highlighted, for example,
in the notable features column.

All of the simulations use cloud-in-cell (CIC, see Hockney \& Eastwood 1982)
interpolation and a Holtzman (1989)
CDM transfer function with $5\%$ baryons, except
CDM16, which uses triangular-shaped-cloud (TSC, see Hockney \& Eastwood
1982) interpolation and a Bardeen et al. (1986, hereafter BBKS)
transfer function.

In order to avoid interference between the initial
interparticle
lattice and the particle-mesh grid (see Gelb 1992 chapter 2),
we begin CDM6 with extra soft forces ({\it i.e.} we set the particle shape
to be a linear sphere density profile
with radius $\eta=5$ grid cells, see Gelb 1992 Appendix I; then
we set $\eta=3.5$ grid cells after
the initial lattice disappears.)
For CDM16, with $144^3$ particles, we use
a $288^3$ grid (we use a $420^3$ grid after $a=0.7$).

High force resolution in a cosmologically significant
box ($\gsim 50$ Mpc) is computationally challenging but can lead
to significantly different results compared with low resolution
simulations.   One of our principle goals
is to study the properties and clustering of resolved halos,
so we are forced to compromise mass and force resolution
by using up to 100 Mpc boxes.
Other authors interested in the detailed properties
of halos, and not clustering,
have concentrated their efforts on very small box sizes.
For example,
Warren et al. (1991) used a
tree code to simulate the formation of halos with
very high particle number
(1097921 particles)
and very high force
resolution
(Plummer softening of 5 kpc proper) in a sphere of radius 5 Mpc.
In another work, Dubinski and Carlberg (1991)
studied CDM halo properties using a tree code with $32^3$
particles
in a sphere of radius 2.3 Mpc.  The initial conditions
were generated in a 8 Mpc box.
The authors used an approximate treatment
of tidal fields and a Plummer softening of 1.4 kpc.
In the present paper the goal is to understand properties of
halos evolved in larger boxes
but with mass and force resolution
significantly better than earlier efforts in boxes
exceeding $\sim 50$ Mpc ({\it e.g.} Davis et al. 1985;
White et al. 1987; Carlberg \& Couchman 1989;
Melott 1990; Park 1990).

\vskip .2truein
\centerline{\it 1.2. Halo Identification}

The standard method for identifying halos from the evolved particle
positions is to identify all particles within a given linking
distance of each other (the friends-of-friends or FOF algorithm).
We developed an alternative, novel procedure
that identifies local density maxima in the smoothed, evolved
density field: DENMAX (see Bertschinger \& Gelb 1991; Gelb 1992 chapter 4).
We first compute a static density field $\delta\rho/\rho$
by interpolating the particles onto a grid.  We then move the particles
according to the equation
$${{d{\vec x}}\over{d\tau}}={\vec\nabla}{{\delta\rho}\over{\rho}}~,
\eqno(1.1)$$
using a ficticious time variable $\tau$ with
$\delta\rho/\rho$ held constant throughout the calculation.
This equation describes a viscous fluid subject to a force proportional
to the density gradient, in the limit of large damping.  Every particle
moves toward a density maximum where it comes to rest.  All particles
lying within closed density contour surfaces around a peak are pushed
toward that peak.  After the particles are sufficiently concentrated
at density peaks, the particles are scooped up and their labels are
recorded.  A halo is composed of these particles with their original
positions restored.  The results of DENMAX depend on the degree of
smoothing used to defined the density field $\delta\rho/\rho$.  We
use trilinear (cloud-in-cell) interpolation with a given grid
({\it e.g.}, $512^3$ or smaller for sensitivity tests) to define the
density field.

After identifying halos, we remove the unbound particles, treating each
halo in isolation.  We compute the potential for each particle $i$,
$\phi_i$, due to all $N_h$ members of a halo:
$$\phi_i = {\sum^{N_h}_{j=1 \atop (j\ne i)}}
\phi(r_{ij});~~~r_{ij}=|\vec{r}_j-\vec{r}_i|~.
\eqno(1.2)$$
The potential is computed once and is fixed throughout the calculation.
(For the P$^3$M simulations we simply use the potential
$\phi(r)=-Gm/(r^2+\epsilon^2)^{1/2}$.  For the PM simulations
we generate $\phi(r)$ by integrating a force table generated by
Monte Carlo sampling the PM force between pairs of particles.)
We then iteratively remove unbound particles as follows.
We compute the energy $E_i=(1/2)m|\vec v_i-\vec v_{\rm cm}|^2+\phi_i$
for each particle $i$, where
${\vec v}_{\rm cm}$ is the mean velocity of the bound particles
at any given stage.  We then remove all particles with $E_i>0$.
The procedure is repeated, each time recomputing
$\vec v_{\rm cm}$, until no more particles are removed.
In all of the DENMAX analyses we remove the unbound particles.
We have also identified halos using the FOF algorithm {\it without}
the removal of unbound particles.

\vskip .2truein
\centerline{\it 1.3. Normalization of the Spectrum}

Most of the simulations are analyzed assuming three normalizations of
the initial, linear CDM power spectrum (a free parameter in the theory).
We define the normalization factor $\sigma_8$  using a tophat sphere
of radius $8h^{-1}$ Mpc:
$$\sigma_8^2\equiv {\intop_0^\infty} d^3k P_{\rm lin}(k)W_{\rm TH}^2(kR)~~;~~
R=8h^{-1}~{\rm Mpc}~\eqno(1.3)$$
with the tophat filter defined as
$$W_{\rm TH}(kR) = {3\over{(kR)^3}}(\sin kR - kR \cos kR)
\eqno(1.4)$$
for comoving wavenumber $k$.  The linear power spectrum of density
fluctuations is
$$P_{\rm lin}(k)={
{\lim_{a_i\to 0}}
}a_i^{-2}P(k,a_i)~.\eqno(1.5)$$

To define the CDM power spectrum we use the primordial scale-invariant
spectrum modulated by the transfer function computed by BBKS
or by Holtzman (1989) with 5\% baryon fraction.
The difference between the two is very small except at high wavenumbers.
We normalize the initial spectrum according to equation~(1.3)
with expansion factor $a\equiv 1$ when $\sigma_8=1$.
We then scale the fluctuations to some early time $a_i$
using linear theory, {\it i.e.} $P(k,a_i)=a_i^{2}P(k,a=1)$.

We generally apply linear theory until the largest $|\delta\rho/\rho|$ on
the initial particle grid is unity.  For the $144^3$ particle simulation CDM16,
however, linear theory is applied until the largest 3-dimensional
displacement is 1 mean interparticle spacing, {\it i.e.} $L/N^{1/3}$ for box
size $L$ and $N$ particles.  The Zel'dovich (1970) approximation is
used to get particle positions and velocities at the end of the linear
regime.
The system is then evolved using the N-body code, with particle
positions and velocities recorded
at various expansion factors $a=\sigma_8$.  (By definition, $\sigma_8
\propto a$.)  In most cases we study the models at $\sigma_8=0.5$, 0.7,
and 1.0.  In the literature, for example, $\sigma_8=0.4$ is known as the
$b=2.5$ biased CDM model because of the assumption that galaxy density
fluctuations are 2.5 times the mass density fluctuations.  According
to the linear biasing paradigm, $b=1/\sigma_8$.  We do not adopt the
linear biasing paradigm because we prefer to identify halos in the nonlinear,
evolved mass distribution.  Note that according to our prescription, the
variance of halo numbers in $8h^{-1}$ Mpc spheres does not necessarily
equal $\sigma_8$.
The COBE measurement of microwave background anisotropy imply (for
a scale-invariant spectrum of density perturbations and the standard
CDM transfer function) $\sigma_8\approx1.1$ (Wright et al. 1992;
Efstathiou, Bond, \& White 1992; Adams et al. 1993).

\vskip .3truein
\centerline{\bf 2. DYNAMIC RANGE: CUMULATIVE MASS FRACTIONS}

In this section we discuss distributions of halos using the cumulative mass
fraction (CMF).

\vskip .2truein
\centerline{\it 2.1. The {\rm CMF}}

The CMF is defined by:
$${\rm CMF}({M})=
{1\over N} {\sum_{{M}'={M}}^{\infty}}~{n}({M}')~{M}',\eqno(2.1)$$
where $N$ is the total number of particles in
the simulation, $M$ is the mass (number of particles) of
a halo, and $n(M)$ is the number of halos containing $M$ particles.
By definition, ${\rm CMF}(0)=1$, ${\rm CMF}(\infty)=0$, and ${\rm CMF}(M)$ is
a decreasing function of $M$.  Note that the particle mass for $N$
particles in a cube of comoving size $L$ is
$$m_{\rm part}(N,L)=4.44\times 10^{9}~M_{\odot}~
\left(128^3\over N\right)\left(L\over 51.2~{\rm Mpc}\right)^3~
.\eqno(2.2)$$

The CMF gives the fraction of mass contained in halos more massive
than $M$.  Although the number and masses of large halos can
fluctuate significantly from simulation to simulation, their contribution
to the CMF gets averaged in the sum of equation~(2.1).  The smallest
mass taken is typically 5 or more particles.  The CMF has the advantage
of summarizing in a nondimensional way all information about the mass
function of halos.  However, it has the disadvantage that halo masses
are not easy to compare with observations.  Also, because halos do not
have sharp outer boundaries, the total mass of a given halo is often not
a well-defined quantity.  We address these problems later by applying
a radial cut-off in order to compare with observations.  Here the motives
are purely theoretical in order to understand the effects of finite resolution.

The first issue we study using the CMF is the difference
between halos identified using DENMAX versus FOF.
Fig.~1 shows the cumulative mass fraction
versus mass for halos found in CDM1(128$^3$,51.2,280)
analyzed with DENMAX and FOF($l=0.1$)
and FOF($l=0.2$), where $l$ is the linking parameter in units of
the mean interparticle spacing.  The DENMAX masses include only the
bound particles, while the FOF masses include all of the identified
particles.
The DENMAX CMFs lie between the FOF CMFs for $l=0.1$ and $l=0.2$.
A smaller FOF linking parameter leads to smaller halos, but also
to a smaller fraction of particles in halos.  The reason for this
is that FOF includes only particles such that the local over-density
exceeds $\sim 2l^{-3}$.  DENMAX, however, gathers all particles around
a peak, even those at lower density.  FOF with $l=0.1$ dissolves
low-density halos.  If $l$ is increased, then FOF merges halos together,
increasing the maximum masses, even when the halos have distinct
substructure (see Gelb 1992 chapter 4, and Fig.~18 below).  DENMAX avoids this
problem: basically, any density concentration visible graphically will
be found by DENMAX.  (Indeed, graphical tests were first used to establish
and test the algorithm.)  Note that more than half of the particles are
associated with {\it some} DENMAX halo, even at early times.  This is the
natural outcome of gravitational instability in a model with small-scale
structure.  Contrary to some expectations, most of the cold dark matter
is not smoothly distributed.

Although the differences in the CMF obtained using DENMAX and FOF are
large, total halo masses are not measured in practice.  It remains to be
seen whether or not observable differences between DENMAX and FOF halos
are large, and whether the results depend on the DENMAX grid or on $l$.
DENMAX has a limitation stemming from the arbitrary choice of a density
grid ($512^3$ for most of the analysis) or equivalently a smoothing scale
for defining the density field.  (Similarly, FOF has its own arbitrary
parameter, $l$.)  We explore these issues later.  For now, our
prejudice is to favor DENMAX because it does not suffer from the obvious
defects of FOF, the dissolving of low-density halos and the merging of halos in
high-density regions.  We include FOF analysis only for comparison with
DENMAX because many authors use FOF ({\it e.g.} White et al. 1987; Carlberg,
Couchman, \& Thomas 1990; Brainerd \& Villumsen 1992).

The lower panel in Fig.~1 shows the effect of the removal
of unbound particles.
The unbinding process systematically reduces
the mass of the halos over the full range of masses, although the effects
are largest for small masses.
We find that the DENMAX results without the removal of
unbound particles are in better agreement with $l=0.2$ FOF.
However, the agreement is not exact; we show later that
FOF occasionally links together visually distinct halos.
Moreover, unbound particles are temporary members of the
halos and therefore should not be included.

Is there a significant simulation--to--simulation variation in the CMF?
In Fig.~2 we show cumulative mass fractions for five simulations.
They are all $128^3$ particle PM simulations
(with $R_{1/2}=280$ kpc comoving) computed
in 51.2 Mpc boxes using
different initial random numbers.
There is very little scatter
at the low mass end and there is considerable scatter at the high mass end.
The fluctuations at the high mass end are due to small number statistics
in these small volumes.
We conclude that the CMF is not sensitive
to simulation--to--simulation fluctuations except
for rare massive halos.

The next important issue is the effect of varying mass
resolution and force resolution.
In Fig.~3 we attempt to determine these effects
by comparing four simulations in 51.2 Mpc boxes
which use initial
conditions taken from an equivalent set
of initial random numbers.
(The same values are used for the initial Fourier transform of
the density fluctuation field for all wavenumbers up to the Nyquist
frequency for each cube.  Thus, the initial conditions for $N=128^3$
are identical to those for $N=64^3$ except that extra high-frequency
power is present with the larger number of particles.)
Mass and force resolution variations cause several effects that we
systematically separate out as we proceed.

The $N=64^3$, $R_{1/2}=560$ kpc comoving PM simulation fails to match
up with the other simulations --- this is not surprising considering
that the force softening is so poor, larger than the size of many halos.
The two very different simulations (the P$^3$M simulation with $64^3$
particles and
$R_{1/2}=52$ kpc comoving versus the PM simulation with $128^3$ particles and
$R_{1/2}=280$ kpc comoving) surprisingly yield very similar
CMFs, but the harder forces in the
P$^3$M simulation actually give rise to halos with higher
circular velocities, an important effect that is discussed
in $\S$ 4. (We show as we proceed that the similarity of the CMF for
these two simulations occurs because increased mass resolution
and increased force resolution both increase the CMF.)
The $256^3$ particle simulation lies above the
others due to the increase in mass resolution
and the presence of more small-scale
power in the initial conditions.

\vskip .2truein
\centerline{\it 2.2. DENMAX Resolution and Box Size}

We need to understand what happens if
we vary the DENMAX grid when analyzing the same simulation.
In Fig.~4 we show the results of several DENMAX
analyses of the P$^3$M simulation CDM12($64^3$,51.2,52)
at $\sigma_8=0.5$.
We see that the DENMAX grid significantly influences
the CMF.  This variation is analogous to the variation of the
CMF with linking length $l$ for the FOF algorithm (cf. Fig.~1).
We demonstrate later, however,
that the circular velocities of the halos
are less sensitive to the DENMAX grid---
this is because circular velocities involve using a cut-off distance
from the local density maximum.  One effect arising from
different DENMAX grids is the inclusion
of distant particles into the halos.
We demonstrate later that the DENMAX
grid influences the break-up of massive halos when the
grids are coarser than the force resolution of the simulation itself.
Because of the density grid sensitivity of DENMAX,
particularly for the total
number of bound particles, we must compare
the CMF from different simulations
using the same effective DENMAX resolution.

Are there significant differences
in the CMF computed in boxes larger than 51.2 Mpc?
In Fig.~5 we show the CMF for two simulations
computed in larger boxes (102.4 Mpc and 100 Mpc).
Since we also use a $512^3$ DENMAX
grid for these simulations, the DENMAX
resolution is only roughly half the resolution of
the 51.2 Mpc simulations analyzed with a $512^3$ DENMAX grid.
The difference is significant (cf. Kundi\'{c} 1991).
In order to separate out the effects due to larger waves in
the initial conditions for the 100 Mpc boxes, compared with
51.2 Mpc boxes,
we compare CDM16($144^3$,100,85)
analyzed with a $512^3$ DENMAX grid with CDM12($64^3$,51.2,52)
analyzed with a $256^3$ DENMAX grid.  (This is done
at $\sigma_8=0.5$ only.)  The two simulations, CDM16 and CDM12, have comparable
force resolution
($R_{1/2}=$ 85 kpc comoving and 52 kpc
comoving respectively) and comparable mass resolution
($m_{\rm part}=~2.3\times 10^{10}~{M}_\odot$ and $3.5\times 10^{10}~{
M}_\odot$ respectively).  The nearly perfect agreement
between CDM16 (100 Mpc box)
analyzed with 512$^3$ DENMAX and CDM12 (51.2 Mpc box)
analyzed
with 256$^3$ DENMAX, and the fact that the two simulations
have comparable force and mass resolution, indicates that
longer waves in the initial conditions do not significantly
affect halo formation. (However, some of the longer waves
have not gone nonlinear yet at $\sigma_8=0.5$.)  This is encouraging
because it means we can use
the simulations in 51.2 Mpc boxes to understand halo properties.
We will discover in Paper II, however, that
the velocity dispersion of pairs of halos
is significantly influenced by the different box sizes.

To quantify the sensitivity of the CMF to DENMAX resolution,
we measure the mass where the CMF equals
20\%, denoted as $M_{20}$.  We
choose 20\% because larger values are not well spanned by
the various simulations and smaller values are more sensitive
to the simulation--to--simulation variations of the massive halos.
We compute, for CDM12($64^3$,51.2,52)
at $\sigma_8=0.5$, the logarithmic slope
$\Delta\log M_{20}/\Delta\log D$ where $D$ is the DENMAX grid spacing.
In Fig.~4, comparing a $512^3$ DENMAX with
a $256^3$ DENMAX analysis, we estimate
$\Delta \log_{10}M_{20}\approx 12.65-13.33=-0.68$ and
$\Delta \log_{10}D=\log_{10}(1/2)$ so $\Delta\log M_{20}/\Delta\log D\approx
2.27$.
Increasing $D$ decreases the DENMAX resolution, thereby
increasing the CMF.
This is because a coarser DENMAX grid tends to pick out larger
masses, {\it i.e.} it cannot resolve substructure.
Comparing a $256^3$ DENMAX grid with a $128^3$ DENMAX grid
we find $\Delta\log M_{20}/\Delta\log D\approx 1.13$.
Comparing a $128^3$ DENMAX grid with a $64^3$ DENMAX grid
we find $\Delta\log M_{20}/\Delta\log D\approx 0.57$.
We therefore see evidence for increasing amounts of substructure
on smaller scales.
Qualitatively similar behavior occurs with the FOF
algorithm (cf. Fig. 1), where the linking parameter plays the role of
the resolution scale.
We demonstrate later that if we impose a radial cut on the DENMAX halos,
as we do when we study circular velocities, the results are not as
sensitive to resolution.

\vskip .2truein
\centerline{\it 2.3. Small-scale Waves}

Fig.~6 is important for understanding the effect of
varying the number of particles---particularly for separating out
the fact that increasing the particle number not only increases
the mass resolution, but it also probes smaller fluctuations in the
initial power spectrum because of the higher Nyquist wavenumber cut-off.
In a discrete system with $N$ particles, the highest wavenumber
represented, is $(2\pi/L)(N^{1/3}/2)$ in each dimension.
We show the results of
a $512^3$ DENMAX analysis from three $R_{1/2}=280$ kpc comoving
PM simulations in 51.2 Mpc boxes which use equivalent initial conditions.
The results are shown at three epochs for 128$^3$ and $64^3$ particles.
We also ran a simulation (CDM9) using 128$^3$ particles, but
the initial conditions are generated by interpolating the
$64^3$ particle case to $128^3$ particles.  Therefore, this
simulation has the same mass resolution as the non-interpolated
128$^3$ particle
simulation (CDM1) but does not have the small-scale waves
present in the non-interpolated simulation.

Apart from the obvious increase in the CMF
due to an increase in mass resolution (explored in greater detail below),
we see in Fig.~6 the effect of the small-scale waves in the initial
conditions --- the non-interpolated
$128^3$ particle simulation has a higher value of the
CMF at small mass relative to the interpolated $128^3$ particle case but
not by much.
Little, Weinberg, \& Park (1991) studied the effect of the removal
of high frequency waves in scale-free models.
Using a PM simulation with $128^3$ particles and $P(k)\propto k^{-1}$,
they found that the nonlinear power spectrum in a simulation
with initial power
above $kL/(2\pi)=32$ set to zero compared very well with
the nonlinear power spectrum in a
simulation with initial power above $kL/(2\pi)=64$ set to zero.
Only small differences appeared on small scales, but
further reductions in the initial cut-off frequency did produce large effects.

\vskip .2truein
\centerline{\it 2.4. Separation of Effects}

We now separate out
the effects of mass and force resolution bearing in mind
that 1) we need to compare simulations in
boxes of different sizes with the same effective DENMAX resolution;
2) the differences in the CMF arising from the inclusion of extra
high and low frequency waves in the initial conditions are small; and 3)
the simulation--to--simulation ({\it i.e.} different
initial random numbers) differences in the CMF are small
below about $10^{13}~{M}_\odot$.
To separate out effects of resolution we
re-examine Figs.~3 through 6.

We first demonstrate that higher mass resolution increases the CMF.
If we examine Fig.~6 we see that the CMF is higher for
the $N=128^3$ particle simulation than for the $N=64^3$
particle simulation using the same force resolution ($R_{1/2}=280$ kpc comoving
in a 51.2 Mpc box).
Comparing $128^3$ and $64^3$ particle simulations, we find
$\Delta\log M_{20}/\Delta\log m_{\rm part}\approx-0.56$.
The minus sign reflects the fact that if the particle mass
increases, then
the mass resolution $\propto 1/m_{\rm part}$
decreases, and therefore $M_{20}$ (or equivalently the
CMF) decreases.
The higher mass resolution simulations
lead to a higher value of the CMF independent
of force resolution.  We also see this
in Fig.~3 by comparing the $256^3$ particle simulation
($R_{1/2}=190$ kpc comoving in a 51.2 Mpc box) with
the $128^3$ particle simulation ($R_{1/2}=280$ kpc comoving
in a 51.2 Mpc box).
The difference between $R_{1/2}=190$ kpc comoving and
$R_{1/2}=280$ comoving
is shown later to have a nontrivial effect on the CMF.

For the $256^3$ particle simulation we find $\log_{10}M_{20}\approx 13.05$
and for the $128^3$ particle simulation we find $\log_{10}M_{20}\approx
12.74$.
Therefore, we find $\Delta\log M_{20}/\Delta\log m_{\rm part}\approx-0.34$.
The effect on the CMF (logarithmic slope)
is smaller going from $256^3$ to $128^3$ particles ($-0.34$) compared
with going from $128^3$ to $64^3$ particles ($-0.56$), but
it is not obvious if and when the results will converge.

The fact that increased mass resolution continues
to increase the CMF in the above comparisons
warrants further investigation.  Is this result still true
when we impose a distance cut from the density peak?
We re-analyze
the three PM simulations
($64^3$, $128^3$, and $256^3$ particles) at $\sigma_8=0.5$
imposing a distance cut of $300$ kpc comoving from
the density peak.  The results
are shown in the top panel of Fig.~7.
In all three cases we do not remove the unbound particles
from halos with raw masses (no cut in radius and no unbinding)
exceeding $1.1\times 10^{13}{M}_\odot$ (location
of vertical line; {\it the transition mass})
to be consistent with the analysis of
the $256^3$ particle simulation.  (In all the analyses of
the $256^3$ particle simulation CDM6
we do not remove the unbound particles from the massive
halos, $M\ge 1.1\times 10^{13}{M}_\odot$, because it
is computationally prohibitive.)
However, the unbinding of the
massive halos has a small effect on the CMF
below the transition mass. To see
this, we show in
the middle panel of Fig.~7 at $\sigma_8=0.5$ the
CMF from the $64^3$ particle simulation
and from the $128^3$ particle simulation
with and without the unbinding of the massive halos.
The effect is negligible
just below the transition mass,
and there is a slight increase in the CMF above
the transition mass.

By examining the top panel of Fig.~7
we find, for the $128^3$ particle
simulation versus the $64^3$ particle simulation,
that $\Delta\log M_{20}/\Delta\log m_{\rm part}\approx-0.23$.  This is less
than $-0.56$, the result when we do not impose a cut of 300
kpc comoving from the density peak.  The CMF itself changes
considerably when we impose a distance cut from the density
peak.  However, we may adopt the position that particles
at such great distances from the center of the halo should
not be associated with estimated measurements of the mass
of observed galaxy halos.  The observed mass of individual
galaxy halos at great distances,
as opposed to dynamical properties inferred by the motions
of stars and gas at small distances, is highly uncertain.
By comparing the $256^3$ particle
simulation with the $128^3$ particle simulation we find
$\Delta\log M_{20}/\Delta\log m_{\rm part}\approx -0.22$.  Again this is less
than $-0.34$, the result when we do not impose a cut of 300 kpc comoving
from the density peak.

Provided that we apply a cut in radius from the density peak,
as we do when
we characterize the halos by their circular velocities in
the next sections, we see that the CMF is less sensitive
to variations in mass resolution than when we do not
impose a cut.  We still do not see a convergence
of the CMF with increasing mass resolution in Fig. 7ab.
However, the $64^3$ particle simulation and the $128^3$ particle
simulation both use $R_{1/2}=280$ kpc comoving.
The $256^3$ particle simulation uses
$R_{1/2}=190$ kpc comoving.  So next we correct for the difference
in force resolution, but first
we demonstrate that higher force resolution also increases the CMF.

In order to see the effect of force resolution we compare simulations
with similar mass resolution.  In Fig.~5 we
compare
CDM12($64^3$,51.2,52)
analyzed with
a $256^3$ DENMAX grid and
CDM11($128^3$,102.4,560)
analyzed with a $512^3$ DENMAX grid.
The mass resolution and the DENMAX
grid resolution are equivalent since the CDM11 box
has eight times the volume of the CDM12 box.  We
see that the higher force resolution simulation yields
a higher value of the CMF.
We find that $M_{20}\approx$ 13.33 for the high resolution
simulation and 13.08 for the low resolution simulation.
If we characterize the force resolution by
$R_{1/2}$, then we find
that $\Delta\log M_{20}/\Delta\log R_{1/2}\approx -0.24$.
This number should be treated with caution
since we are comparing simulations with $R_{1/2}=52$ kpc comoving
versus $R_{1/2}=560$ kpc comoving---this is a wide range and
DENMAX behaves unreliably in very low resolution simulations.
We do not
have two P$^3$M CDM simulations with comparable mass resolution
but with significantly different Plummer softenings.

The increase in the CMF for higher force resolution
simulations is verified by comparing
CDM12($64^3$,51.2,52)
with CDM8($64^3$,51.2,560)
in Fig.~3,
but again the force resolution in CDM8
is extremely poor.

As a final comparison of force resolution effects, we compare
CDM12($64^3$,51.2,52)
with CDM7($64^3$,51.2,280)
both analyzed with a $512^3$ DENMAX grid.
We find $\Delta\log M_{20}/\Delta\log R_{1/2}\approx
-0.56$.  The range of force softenings in this comparison is still
large but at least $R_{1/2}=280$ kpc comoving is more reasonable than
560 kpc comoving.
In subsequent sections we compare the halos characterized
by their circular velocities and particular attention
is paid to force resolution comparing results for
PM versus P$^3$M simulations.  So we return to force resolution
then.

As a final test of the convergence of the CMF
with increasing mass resolution, we first use the above force resolution
analysis to estimate the effect on the CMF from a
$R_{1/2}=280$ kpc comoving PM simulation versus a $R_{1/2}=190$
kpc comoving PM simulation.
To do so, we compare the
$64^3$ particle, $R_{1/2}=52$ kpc comoving P$^3$M simulation
with the
$64^3$ particle, $R_{1/2}=280$ kpc comoving PM simulation
(both computed
in 51.2 Mpc boxes and analyzed with a $512^3$ DENMAX grid)
imposing a 300 kpc comoving
cut from the density peak of the halos.
The logarithmic slope is $\Delta\log M_{20}/\Delta\log R_{1/2}-0.74$.
If we multiply $-0.74$ by
$\Delta \log_{10}R_{1/2}=-0.18$, {\it i.e.} the difference between
the softening of the $R_{1/2}=190$
kpc comoving PM simulation and the $R_{1/2}=280$ kpc comoving
PM simulation,
we get $\Delta \log_{10}M_{20}\approx 0.13$.
Therefore, we can estimate that the $256^3$ particle $R_{1/2}=190$
kpc comoving PM
simulation (with a 300 kpc comoving distance cut)
would have $\log_{10}M_{20}\approx 11.91-0.13=11.78$
if it were computed using a $R_{1/2}=280$ kpc comoving
PM simulation.

Now if we compare the re-scaled (to $R_{1/2}=280$ kpc comoving)
$256^3$ particle result
with the $R_{1/2}=280$ kpc comoving $128^3$ particle
PM simulation, all
with a 300 kpc comoving distance cut,
we get $\Delta\log M_{20}/\Delta\log m_{\rm part}\approx
-0.11$ compared with the old value of $-0.22$.
This is encouraging because this logarithmic slope,
$-0.11$, is still better
than the logarithmic slope $-0.23$ computed earlier by
comparing a $128^3$ particle simulation with
a $64^3$ particle simulation.
Graphically (as depicted in Fig.~7c)
this corresponds to moving
the CMF for the $256^3$ particle simulation in the top
panel of Fig.~6 $0.13$ units
to the left.

We now see that the agreement between the $128^3$ particle case
and the $256^3$ particle case is much better (Fig. 7c).  There is still
a slight increase in the CMF on small mass scales.  However,
this is consistent
with the fact that the $256^3$ particle simulation has more
small scale power in the initial conditions compared with
the $128^3$ particles simulation.  This
effect was demonstrated earlier.

Using simulations analyzed with the same effective DENMAX resolution,
we found the following:
1) Higher mass resolution leads to larger values
of the CMF independent of force resolution.
The effect is smaller when we impose a distance cut
from the density peak of the halos.
The difference between the
$4.4\times 10^{9}{M}_\odot$
and the
$3.5\times 10^{10}{M}_\odot$
simulation (using a distance cut of 300 kpc comoving from the density
peaks of the halos)
is small, $\Delta\log M_{20}/\Delta\log m_{\rm part}\approx -0.23$.
The difference between the
$5.5\times 10^{8}{M}_\odot$ simulation
and the
$4.4\times 10^{9}{M}_\odot$ simulation
(using a 300 kpc comoving cut and correcting
for the difference in force softening)
is $\Delta\log M_{20}/\Delta\log m_{\rm part}\approx
-0.11$.  The difference has decreased in the very high mass
resolution simulation indicating that convergence of the CMF
with mass resolution is plausible.
2) Higher force resolution
leads to larger values of the CMF independent of mass resolution.
We examine the effects on the formation of halos arising
from different force resolution P$^3$M simulations in
subsequent sections.
3) Longer waves in the initial conditions (100 Mpc box versus
a 51.2 Mpc box) do not significantly affect the CMF.
4) Smaller waves in the initial conditions ($64^3$ particle
initial conditions interpolated to $128^3$ particles versus
true $128^3$ particle initial conditions) do not significantly affect the
CMF, aside from a small effect on small mass scales.
5) Larger DENMAX grids better resolve substructure;
this lowers the CMF.  The results are sensitive to the different DENMAX
grids so it is important to compare CMFs using the same
effective DENMAX resolution.  However, we show later
that the results are less sensitive when we compute
circular velocities which are what we use to compare
the simulated halos with the observations.

\vskip .2truein
\centerline{\it 2.5. Press-Schechter Theory}

As a final application of the CMF, we compare the simulations
with the predictions of the Press-Schechter theory
(Press \& Schechter 1974).
The Press-Schechter formalism estimates the fraction of mass in
bound halos with masses $>M$ to be the fraction of the mass
whose linear density, averaged over a scale $M$, exceeds $\delta_c$:
$$P(M)={\rm erfc}
\left[ {\delta_c \over 2^{1/2}\sigma_0(M) } \right]~,
\eqno(2.3)$$
where erfc is the complementary error function.  One may regard
$\delta_c$ as a free parameter, although it is often taken to
equal the critical over-density for uniform spherical collapse
in an Einstein-de Sitter universe, $\delta_c=1.68$.
The rms density $\sigma_0(M)$ is computed from the linear
power spectrum, smoothed with an appropriate filter (window function).
We use either a Gaussian window function, $W(k,R_f)=\exp(-0.5x^2)$,
or a tophat window function, $W(k,R_f)=3(\sin x-x\cos x)/x^3$, where in
both cases $x\equiv kR_f$.  The generalized spectral moments (to be used
below) are defined as follows:
$$\sigma_n^2(M)
\equiv{\intop_0^\infty}
{4\pi k^2 P(k)\,W^2(k,R_f)}
{k^{2n}}
dk~.\eqno(2.4)$$
For a Gaussian window function, the smoothing radius $R_f$ is related
to the mass as follows: $M=(2\pi)^{3/2}\rho_0 R_f^3$.  For a tophat
window function, $M=(4\pi/3)R_f^3$.

Press \& Schechter (1974) estimated the mass function of bound halos
as $n(M)d\ln M=2\rho_0 (dP/dM) d\ln M$, where
$\rho_0$ is the comoving background mass density.  The factor
of two is needed for normalization, but has since been derived
analytically by Bond et al. (1991).  The final result is:
$$n(M)d\ln M=
\left(2\over \pi\right)^{1/2}
{\rho_0\over M}\,
{\delta_c\over{\sigma_0(M)}}\,
\left\vert d\ln\sigma_0\over d\ln M\right\vert\,
\exp{\left[{ -{1\over 2} {\left({ {\delta_c\over
\sigma_0(M)} }\right)}^2
}\right]}
d\ln M~.\eqno(2.5)$$
We convert equation~(2.5) into a CMF using
$${\rm CMF}(M)={1\over\rho_0}\intop_{\ln M}^\infty n(M') M' d\ln M'~.
\eqno(2.6)$$

We evaluate equation~(2.6) using $\sigma_8=0.5$, 0.7, and 1.0
linear normalizations of the BBKS CDM power spectrum.
We try $\delta_c=1.44$ ({\it e.g.} Carlberg \& Couchman 1989),
$\delta_c=1.68$ ({\it e.g.} Efstathiou et al. 1988;
Brainerd \& Villumsen 1992), and $\delta_c=2.0$, for both a gaussian and a
tophat window function.
Theoretical predictions of the Press-Schechter theory are compared with
CMFs measured from the high resolution N-body simulations
CDM12($64^3$,51.2,52; particle mass $3.5\times10^{10}{M}_\odot$) and
CDM16($144^3$,100,85; particle mass $2.3\times10^{10}{M}_\odot$) in
Fig.~8.

First we consider the halos identified according to the
FOF algorithm with a linking parameter $l=0.2$.
Fig.~8a shows that the two simulations, at three different epochs,
yield reasonably good agreement with the Press-Schechter predictions for
a tophat window function with $\delta_c=2.0$.  Only slightly worse
agreement obtains with a gaussian window function with $\delta_c=1.68$.
Note that the simulated mass distributions are broader than predicted.
The high mass tails of the distributions actually match very well the
Press-Schechter predictions for a tophat window function with $\delta_c=
1.68$, but there are fewer low mass halos than predicted.  Evidently
this is because they are subsumed into more massive halos, at least
with the FOF recipe, with greater efficiency than implied by the analytical
model.  Although the agreement with the Press-Schechter theory is not perfect,
the errors do not grow with epoch; the analytical theory appears to give the
correct scaling of masses as the clustering strength increases.  Our result
here differs from that of Brainerd \& Villumsen (1992), who found the
departures growing as clustering progresses.

Fig.~8b shows similar results for a FOF linking length
$l=0.1$.  Now $\delta_c$ must be increased (from 1.68 to 2.0 for the
gaussian window function) to account for the smaller masses of the halos
defined at a higher over-density.  However, the agreement at small masses
is significantly worsened.

Fig.~8c shows CMFs for CDM12($64^3$,51.2,52)
computed using DENMAX, compared with
Press-Schechter theory for a gaussian window function with $\delta_c=2.0$.
The top set of data points (filled circles and solid curves) are for raw
DENMAX masses, with no removal of unbound particles (which would decrease
the CMF by about 10\%) and with no radial cut.  The bottom set (dashed
curves and crosses) have excluded unbound particles and those beyond a
comoving radius of 200 kpc from the peak.  There are several important
things to notice.  First, at early epochs, the raw DENMAX CMF agrees
well with the Press-Schechter theory.  At high masses the DENMAX
distributions are similar to those obtained using FOF with $l=0.1$ while at
low masses they match the $l=0.2$ case better.  DENMAX breaks up the more
massive clumps found with $l=0.2$ while preserving the subclumps as
individual halos.

Second, as clustering increases, the CMF grows less rapidly than the
Press-Schechter prediction.  This effect appears to be due to the ability
of DENMAX to find substructure in halos merged by FOF.  Thus, although we
disagree with Brainerd \& Villumsen (1992) about the results from
FOF, we agree that the actual halo mass distribution grows
less rapidly than predicted by Press-Schechter theory.  The agreement
could be improved if $\delta_c$ were to grow with epoch.  In fact, at
very early epochs (when there are fewer than 100 particles per group) the
fit to the simulations is good with a gaussian window and $\delta_c=1.68$.

The third point to note from Fig.~8c is that the radial truncation
of the halos makes a big difference in the masses.  Thus, the halos are
very extended, a point that we will demonstrate more clearly later.

In summary, halo mass functions depend on how the halos are defined.
Earlier workers ({\it e.g.}, Efstathiou et al. 1988; Carlberg \& Couchman 1989)
found good agreement between the Press-Schechter theory and simulations.
However, the simulations were analyzed with a low resolution group finder,
FOF ($l$=0.2), and the halos contained relatively few particles.  Our results
agree with this work, but show further that the Press-Schechter theory does
not match well the CMF when higher resolution is used to identify halos
made of thousands of particles.  The disagreement is in the sense that
the simulated halos are less massive than predicted.  This occurs not
because large halos have failed to collapse.  Rather, merging does not
immediately erase the substructure in large halos, contrary to the
assumptions made in the Press-Schechter theory.

\vskip .3truein
\centerline{\bf 3. DISTRIBUTIONS OF HALOS: BACKGROUND}

\vskip .2truein
\centerline{\it 3.1. The Schechter Luminosity Function}

We need to
define physically motivated catalogs of halos in order to
understand further the effects of dynamic range on halo formation
and in order to compare the simulations with the observations.
Total bound mass, as in the previous section, is only
one way to characterize the halos.  We can also ask
how much mass is contained within a specified radius.
This is equivalent to specifying $V_{\rm circ}=(GM/R)^{1/2}$.
Empirically, $V_{\rm circ}$ is found to be nearly
independent of $R$ and to correlate well with
optical luminosity.  We will use these correlations---
the Tully--Fisher (1977) relationship for spiral
galaxies and the Faber--Jackson (1976) relationship for elliptical
galaxies---
to assign a luminosity to each halo.

Observations of spiral galaxies
are measured in terms of their circular velocity and
observations of elliptical galaxies are measured in
terms of their average central radial
velocity dispersion.
(Technically, the elliptical observations are luminosity weighted measurements
of radial velocities along the line-of-sight.)

We realize that we cannot adequately relate internal velocity dispersions
of dark matter to velocity dispersions of centrally concentrated stars.
Nevertheless, we define the quantities
$\sigma_1$ and $\sigma_r$ ($\sigma_r$ is closer to what
the observers measure) from the simulated galaxy halos as
follows:

$$\sigma_1^2(R)=
{1\over{3N_c}}
{\sum_{i=1}^{N_c}}
{\left|{  {\vec v_i-\vec v_{\rm cm}} }\right|}^2
\ ,\quad
\sigma_r^2(R)=
{1\over{N_c}}
{\sum_{i=1}^{N_c}}
{\left|{
{ {\left({\vec v_i-\vec v_{\rm cm}}\right)} \cdot {\hat{\vec r}_i} }
}\right|}^2
\ ,\eqno(3.1)$$

{\parindent 0pt
where $N_c$ represents the number of bound particles within a distance
$R$ from the local density
maximum and ${\hat{\vec r}}_i$ is the unit vector from the local
density maximum to particle $i$.
We do not attempt to distinguish
the simulated halos as spirals or ellipticals; rather,
we characterize all of the simulated halos in terms of their
circular velocities.}

Because the velocity dispersion tensor is radially anisotropic we find
that $\sigma_1$ is typically $\sim 20\%$ lower than $\sigma_r$.
We study both quantities, using various cut-off radii (typically a few
hundred kpc comoving),
when comparing the velocity dispersions of massive simulated halos
(perhaps associated with elliptical galaxies) with observations
of the velocity dispersions of centrally concentrated stars.
In order to test if either $\sigma_1$ or $\sigma_r$ is a useful
statistic, and because the stars are in orbits
with smaller apapses than the dark matter, we use a crude, linear scaling
law (derived from observations of M87) as
discussed in greater detail in  $\S~4.4$.

For $\Omega=1$ and $h=1/2$, the circular velocity, for
an assumed spherical halo, as
a function of total particle number in the simulation, $N$,
and the comoving box size in Mpc, $L$, is:
$$V_{\rm circ}(R)=7.97 \kms \sqrt{N_c(R)}\left(128^3\over N\right)^{1/2}
\left(L\over 51.2~{\rm Mpc}\right)^{3/2}
\left(300~{\rm kpc} \over {Ra/a_0}\right)^{1/2}~,\eqno(3.2)$$
where $N_c(R)$ is the total number of bound particles within a
comoving distance $R$ from
the smoothed density maximum found by DENMAX.
The present epoch is $a=a_0\equiv\sigma_8$.  In most of the figures,
we assume that $a=a_0$ and we consider different possible normalizations
by varying $\sigma_8=a_0$.  In one case below (Fig. 16), we fix $a_0$
and look at the evolution of halos for different $a$.  In all cases,
we take $R$ to be a comoving radius ({\it i.e.}, a proper radius at $a=a_0$)
and we use the proper radius $Ra/a_0$ in the denominator.  To get
circular velocities measured at a fixed comoving radius, we set $a=a_0$.

In order to compute the observed distribution
of galaxies as a function of
$V_{\rm circ}$, {\it i.e.} $N(V_{\rm circ})\Delta V_{\rm circ}$,
we assume a Schechter (1976) luminosity function with the form
$$\Phi({\cal L})d{\cal L}=\Phi^{*}\exp{\left(-{\cal L}/{\cal L}_{*}\right)}
{\left({\cal L}/
{\cal L}_{*}\right)}^\alpha
d{\left({\cal L}/{\cal L}_{*}\right)}~,\eqno(3.3)$$ where $\Phi({\cal L})
d{\cal L}$ is
the density of galaxies in the luminosity range ${\cal L}$
to ${\cal L}+d{\cal L}$.
We convert equation~(3.3) into counts of halos
in a $(51.2~{\rm Mpc})^3$ comoving volume as a standard reference
for all of the simulations in bins of $V_{\rm circ}$
using a relationship for ${\cal L}={\cal L}(V_{\rm circ})$.
We also use
blue magnitudes and selected values of $\Phi^{*}$ and
${\cal L}^{*}_{B_T}$ (both assuming $h=1/2$), and a value of $\alpha$.

We define the distribution of halos, or number of
halos binned by $V_{\rm circ}$, as
$$N(V_{\rm circ})\Delta V_{\rm circ}=\left(51.2~{\rm Mpc}\over L\right)^3
\widetilde{N}(V_{\rm circ})\Delta V_{\rm circ}~,\eqno(3.4)$$
where $\widetilde{N}(V_{\rm circ})$ is the number of halos found in
the simulation with
circular velocities in the range $V_{\rm circ}\pm\Delta V_{\rm circ}/2$
with $\Delta V_{\rm circ}=25\kms$.  The factor
$(51.2~{\rm Mpc}/L)^3$ is used to scale all of the results
to comoving volumes $(51.2~{\rm Mpc})^3$ for comparison.

We compute the corresponding mean number of galaxies
from the observations as follows, assuming ${\cal L}$
is related to $M_{B_T}$ and $M_{B_T}=f(V_{\rm circ})$
for some function $f$ given below:
\vbox{
$$N_{\rm Schechter}(V_{\rm circ})\Delta V_{\rm circ}=
(51.2~{\rm Mpc})^3
{\intop_ {x^{(1)}}       ^{x^{(2)}} }\Phi(x)~dx$$

$$M^{(1)}_{B_T}=f(V_{\rm circ}+\Delta V_{\rm circ}/2)$$

$$M^{(2)}_{B_T}=f(V_{\rm circ}-\Delta V_{\rm circ}/2)$$

$$\Phi(x[M_{B_T}])=\Phi^*x^\alpha \exp(-x)~;~x\equiv
10^{(M^*_{B_T}-M_{B_T})/2.5}={{\cal L}\over {\cal  L}^*}~.\eqno(3.5)$$
}

{\parindent 0pt We use the central values of parameters
found by Efstathiou, Ellis, \& Peterson (1988):
$\Phi^{*}=(1.56\pm 0.34)\times 10^{-2}h^3~{\rm Mpc}^{-3}$,
$M^{*}_{B_T}=-19.68\pm 0.10-2.5\log_{10}h^{-2}$,
and $\alpha=-1.07\pm0.05$.}

For the function $f(V_{\rm circ})$ for spiral galaxies
we use the blue Tully-Fisher relation
from Pierce \& Tully (1988):
$$f_{\rm spiral}(V_{\rm circ})\equiv M_{B_T}=-6.86\log_{10}(2V_{\rm
circ})-2.27+5\log_{10}(50/85)+0.569~.\eqno(3.6)$$
The term $5\log_{10}(50/85)$ is used to convert from
a Hubble constant of 85 $\kms~{\rm Mpc}^{-1}$ to
50 $\kms~{\rm Mpc}^{-1}$.  The term 0.569
is used to correct for random inclinations following
Tully \& Fouque (1985).

For the function $f(V_{\rm circ})$ for elliptical galaxies
we use the Faber-Jackson relation
from our fit (unpublished)
to elliptical data of Faber et al. (1989),
assuming a Hubble constant of 50 $\kms~{\rm Mpc}^{-1}$:
$$f_{\rm elliptical}(V_{\rm circ})\equiv
M_{B_T}=-6.6364\log_{10}(\sigma_1)-5.884~,\eqno(3.7)$$
where we relate $\sigma_1$ to $V_{\rm circ}$ using:
$$\sigma_1=F {V_{\rm circ}\over \sqrt{3}}~.\eqno(3.8)$$
The factor $F$, discussed in the next section,
is estimated from the simulations.  This use of $\sigma_1$, however,
is an oversimplification (mostly affecting
high $V_{\rm circ}$) for reasons discussed earlier.
Again, we re-examine the high mass halos in detail
in $\S~4.4$, where we use $\sigma_1$, $\sigma_r$, and a linear scaling
law derived from M87.

The final ingredient is to
assume that 70\% of the galaxies are spirals
and 30\% are ellipticals.  This
is also the assumption used by Frenk et al. (1988).
In other words, we
add together the results for spirals using
equation~(3.6) to relate circular velocities to
absolute magnitudes and weighting equation~(3.5)
by 0.7 with
the results for ellipticals
using equation~(3.7) to relate circular velocities to
absolute magnitudes and weighting equation~(3.5)
by 0.3.
Dressler (1980), however, found a higher concentration of ellipticals in
rich clusters compared with lower density regions.
Postman \& Geller (1984)
found for the CfA survey that 1) the
relative numbers of galaxies are 65\% spirals, 23\% S0's, and 12\% ellipticals
and 2) there is a dramatic increase in the relative number of spirals
in the field compared with dense regions.  These percentages
can alter the
estimates at the high mass end.

\vskip .2truein
\centerline{\it 3.2. $\sigma_1$ versus $V_{\rm circ}$}

The factor $F$ in equation~(3.8) is measured empirically
from the simulation CDM16($144^3$,100,85)
using DENMAX halos analyzed with a $512^3$ grid.
White et al. (1987) used $F=1$ (in our notation)
but the same authors used $F=1.1$ in
Frenk et al. (1988).

Because the Plummer softening in CDM16 is $\epsilon=65$ kpc comoving
(or $R_{1/2}=85$ kpc) we cannot directly determine
$\sigma_1(R)$ or $V_{\rm circ}(R)$ at the distances where
optical observations of real galaxies are made.
Optical observations of central velocity
dispersions of large elliptical galaxies are made on scales
of a few kpc to $\sim 6$ kpc (see Franx, Illingworth, \& Heckman 1989).
Optical observations of circular velocities of large spiral
galaxies are made out to $\sim 10$ kpc.  Rubin et al. (1985)
studied 16 large spiral galaxies where they could measure
velocities out to large radii.  The average maximum distance
for which they made measurements was 16.4 kpc and the maximum
distance for the 16 galaxies was 51.2 kpc.
We consider the limitations arising from our measurements at large
radii as we proceed.

What is a value of $R$ for computing $\sigma_1$
where the results are independent of $R$?
In Fig.~9 (top panel) we show $\sigma_1$ evaluated using
$R=100$ kpc comoving versus $R=200$ kpc comoving.
The slight increase in $\sigma_1$ for $R=200$ kpc comoving versus $R=100$ kpc
comoving (top panel) indicates that contributions from particles at
large separations are still important for the most massive halos.
We find this trend to be larger when comparing results from
$R=50$ kpc comoving versus $R=100$ kpc comoving,
indicating that $R=100$ kpc comoving is too small.
We find this trend to be small when comparing results
from $R=200$ kpc comoving versus
$R=300$ kpc comoving, indicating that $R=200$ kpc comoving
is adequate.
We find similar results at $\sigma_8=0.7$ and $\sigma_8=1.0$.

What is a value of $R$ for computing $V_{\rm circ}$
where the results are independent of $R$?
In Fig.~9 we also show computations for
$V_{\rm circ}$
using $R=100$ kpc comoving,
200 kpc comoving, and 300 kpc comoving.
The results
indicate that $R=200$ kpc comoving is acceptable
(bottom panel) while $R=100$ kpc comoving again
is too small (middle panel).

What is an empirical value of $F$ in equation~(3.8)?
In Fig.~10 we show $V_{\rm circ}/\sigma_1$
versus $V_{\rm circ}$ (all computed with $R=200$ kpc comoving)
at $\sigma_8=0.5$, 0.7, and a=1.0
for halos from CDM16($144^3$,100,85).
The solid lines indicate $F=1$ ({\it i.e.} $V_{\rm circ}/\sigma_1=\sqrt{3}$)
and the dotted lines indicate $F=1.1$ ({\it i.e.} $V_{\rm
circ}/\sigma_1=\sqrt{3}/1.1$).
There is less
scatter for high values of
$V_{\rm circ}$ versus low values of $V_{\rm circ}$.
The factor $F$ affects the conversion of $V_{\rm circ}$
to $\sigma_1$ for ellipticals.  Ellipticals dominate at
the high mass end where,
at $\sigma_8=0.5$,
$F=1.1$ works slightly better than $F=1$.
However,
when we show the observed number of halos in $\S$ 4
we use both $F=1$ and $F=1.1$---the latter yields
fewer bright halos since it effectively raises
$\sigma_1$ for a given $V_{\rm circ}$
implying a brighter elliptical galaxy (or equivalently, $F=1.1$
effectively assigns a smaller stellar velocity dispersion
for a given $\sigma_1$).

In summary, we compute the number of halos
scaled to $(51.2~{\rm Mpc})^3$ comoving volumes assuming
a Schechter luminosity function ($\Phi^{*}=1.56\times 10^{-2}h^3~{\rm
Mpc}^{-3}$
and $M^{*}_{B_T}=-19.68-2.5\log_{10}h^{-2}$ with $h=1/2$
and $\alpha=-1.07~$).
We assume
70\% of the galaxies are spirals with
a Tully-Fisher relation given by equation~(3.6)
and 30\% of the galaxies are ellipticals
with a Faber-Jackson relation given by equation~(3.7).
We convert elliptical measurements
in terms of $\sigma_1$ to $V_{\rm circ}$
using equation~(3.8) for both $F=1$ and $F=1.1$.
For the most massive halos $F=1$ is adequate except at
$\sigma_8=0.5$ where $F=1.1$ is slightly better.
\vfill
\eject
\vskip .2truein
\centerline{\it 3.3. Circular Velocity Profiles}

We now examine circular velocity profiles and
1-dimensional velocity dispersion profiles from
CDM12($64^3$,51.2,52) in Fig.~11.
We extract several facts from these plots.  First, the circular
velocities are very flat for nearly all the halos, except for a
few massive ones, for $R\gsim$ 150 kpc comoving.  We also see
that the circular velocities are flat for many of the midsize
halos down to about 80 kpc comoving  (twice the Plummer softening length).
We conclude that 150 kpc comoving is a good place
to characterize the circular velocities for this simulation
at all epochs.  We are making a significant error only
for a handful of the most massive halos.
The rising circular velocities for the most massive
halos are not an artifact of softening (see $\S~4.4$).

We also study CDM1 ($R_{1/2}$= 280 kpc comoving) and CDM16 ($R_{1/2}=85$
kpc comoving) and we find that $R=300$ kpc comoving
is suitable for CDM1
and $R=200$ kpc comoving is suitable for CDM16.
In most cases, we use $R=150$ kpc comoving for the $R_{1/2}=52$ kpc comoving
simulation; $R=200$ kpc comoving for the $R_{1/2}=85$ kpc comoving
simulation; and $R=300$ kpc comoving for the $R_{1/2}=280$ kpc comoving
simulation.  These values are chosen where $V_{\rm circ}$
is flat for nearly all of the halos.

We compare our circular velocity profiles to simulations by others
with much higher force resolution.
The $\Omega=1$ CDM simulations of Dubinski \& Carlberg (1991)
used 33000 particles in a 2.3 Mpc radius sphere.
The typical particle mass is $1.2\times 10^8{M}_\odot$
and the Plummer softening is 1.4 kpc.  In their figure~4
they show several circular velocity profiles with halos
that have maximum circular velocities of about 290 $\kms$.
These halos have flat circular velocities between about 10 kpc
and 60 kpc (there is a very slight decrease over this range).
The circular velocities rise on a scale a few times the Plummer
softening length as we also find in our simulations.

The simulations of Warren et al. (1991) used roughly
a million particles in a 5 Mpc radius sphere and
a Plummer softening of 5 kpc.  For circular velocity
profiles
that have maximum circular velocities of about $150\kms$
they found that the profiles are rising out to a distance of
about 30 kpc---again,
several Plummer lengths.  They also found that their profiles
are falling typically beyond a distance of about 40 kpc.  In our P$^3$M
simulations we do not find falling
circular velocity profiles until a distance of about 100 kpc.  One
reason for this disrepancy is that Warren et al. did not
not use a
CDM power spectrum---rather, they used $P(k)\propto k$
on large scales with a sharp transition, at 1.5 Mpc,
to $P(k)\propto k^{-2}$ on small scales.  The
behavior of circular velocities is a function of the effective index in
the initial power spectrum (Hoffman \& Shaham 1985).

We assume that if we had used force softening below the
typical $\sim$10 kpc observed scale, as in the simulations
of Dubinski \& Carlberg (1991) and Warren et al. (1991),
that our circular velocity profiles might remain flat down
to these scales.  Therefore, we do not expect to make a significant
error by estimating $V_{\rm circ}$ using $R\gsim$ 150 kpc comoving.
We cannot use circular velocities to characterize
the most massive halos because observational data for
massive halos are based on velocity dispersions, not circular velocities.
We explore the properties of $\sigma_1$ below.

There is cause for concern when using simulations with
force softening far beyond a few kpc---the scale beyond
which most spiral galaxies are observed to have flat rotation curves.
Are we able
to adequately resolve individual halos?
We comment on several issues related to this question.
1) Using $\sim (1/\Phi^{*})^{1/3}$, and $\Phi^{*}=1.95\times 10^{-3}~{\rm
Mpc}^{-3}$, we find the mean spacing between bright
galaxies is $\sim 8$ Mpc.  This is much greater than our fiducial
radius $\sim 200$ kpc.  Observations show flat rotation curves
``as far as the eye can see''
for most spiral galaxies (Rubin et al. 1985).
It is therefore possible that real galaxies have flat
rotation curves beyond $200$ kpc.
2) The mean galaxy spacing is much smaller in rich clusters.
It is possible that some of our massive halos are mergers
where dissipative effects might allow many
galaxies to survive in a single halo
(White \& Rees 1978; Katz \& White 1993).
In Paper II we break up these
systems using various methods in an attempt to estimate the effects
on clustering.
In this paper, however, we consider the massive halos
at face value and we examine the implications for CDM in $\S$ 4.
3) We compare results
from the $\epsilon=40$ kpc comoving simulation with
results from the $\epsilon=65$ kpc comoving simulation and we ask
if the distributions of halos are significantly different.

Last, we show $\sigma_1(R)$ for the $\epsilon=40$ kpc comoving
simulation CDM12($64^3$,51.2,52) in Fig.~12.
(These are the same halos shown in Fig.~11.)
The first thing we notice is that the profiles are very
flat down to about 40 kpc comoving, the Plummer softening scale.
(On smaller scales we are limited by both force and mass resolution.)
Also, the profiles for the most massive halos are flat down
to typically 100 kpc comoving.
In $\S~4.4$ we use $\sigma_r$ computed at large
radii (similar to $\sigma_1$)
to compare simulated halos with the observations.

\vskip .3truein
\centerline{\bf 4. DISTRIBUTIONS OF SIMULATED HALOS}

\vskip .2truein
\centerline{\it 4.1. Overview}

We now study the distributions of simulated halos as a function
of $V_{\rm circ}$.
The results of the computations of $N(V_{\rm circ})\Delta V_{\rm circ}$
scaled to $(51.2~{\rm Mpc})^3$ comoving volumes
are presented
in Figs.~13 through 16.
We include observational estimates using both
$F=1$ and $F=1.1$ to relate $\sigma_1$ to $V_{\rm circ}$ for
comparison.

We focus our efforts on the following three simulations:
CDM1($128^3$,51.2,280),
CDM12($64^3$,51.2,52), and CDM16($144^3$,100,85).
CDM1 offers good mass resolution
($m_{\rm part}=4.4\times 10^{9}~{M}_\odot$), CDM12
offers good force resolution (Plummer softening of 40 kpc comoving),
and CDM16 offers fairly good mass and force resolution
($m_{\rm part}=2.3\times 10^{10}~{M}_\odot$; Plummer softening
of 65 kpc comoving) yet is computed in a 100 Mpc box.
Again, we demonstrated in $\S$ 2 that the CMF is not very sensitive
to the box size but we demonstrate in Paper II
that clustering statistics require boxes larger than 51.2 Mpc
on a side.

Our goal in the following sections
is to attempt to constrain the amplitude of
the primeval density fluctuations of the $\Omega=1$ CDM model
from halo circular velocity distributions.
We devote separate subsections for both high mass and low mass
halos, which require special treatment for determining reliable
simulated and observed distributions.

\vskip .2truein
\centerline{\it 4.2. Circular Velocity Distributions of Simulated Halos}

We measure $N(V_{\rm circ})\Delta V_{\rm circ}$ from the simulations.
We ask the questions:
1) Over which range of
circular velocities do the results agree with the observations?
2) Over which range of circular velocities do the results disagree
with the observations?
3) Do the results depend on numerical resolution and techniques
for identifying halos?

In Fig.~13 we show
$N(V_{\rm circ})\Delta V_{\rm circ}$ at $\sigma_8=0.5$, 0.7 and 1.0 for
an analysis of CDM1($128^3$,51.2,280)
using $512^3$ DENMAX
and FOF($l$=0.1) and FOF($l$=0.2).
We see from Fig.~13 that the number of halos agrees
with the observations very well from about $150\kms$ to
$350\kms$ for DENMAX and FOF($l$=0.2).  The results for
FOF($l$=0.1) do not fare as well.
These statements are true for all three epochs; however,
the excess number of massive halos gets worse with increasing
$\sigma_8$.
DENMAX is a compromise between FOF($l$=0.2)
which sometimes merges halos and FOF($l$=0.1) which fails to produce
some halos.  These results are encouraging for studies that
use FOF($l$=0.2) such as Frenk et al. (1988); however,
FOF($l$=0.2) occasionally links together visually distinct halos.

We now study the effects of force resolution, choice of DENMAX grid,
and choice of $R$ (used to compute $V_{\rm circ}$)
on $N(V_{\rm circ})\Delta V_{\rm circ}$.
We show $N(V_{\rm circ})\Delta V_{\rm circ}$ at $\sigma_8=0.5$
for CDM12($64^3$,51.2,52) in Fig.~14.
The first thing we notice is that
the agreement with the number of simulated halos with
the observations from $150\kms$ to $350\kms$
is even better than it is
for the low force resolution PM simulation discussed above,
particularly for $V_{\rm circ}\sim 200 \kms$.
We also see that the results are not very sensitive to
the choice of $R$ except for the few very massive halos.
This is not surprising since most of the circular velocity
profiles are flat beyond 150 kpc comoving except for the most
massive halos---cf. Fig.~11.

It is encouraging
that the results are not very sensitive to the choice of
DENMAX grid except for the most massive halos and for
the $64^3$ grid.  This is not true for the halo masses
described by the CMF in $\S$ 2---we show later
that this is because the different DENMAX grids
significantly affect peripheral particles
beyond the distance $R$ used to compute the circular velocities.  It
is not surprising that the very coarse $64^3$ grid fails
to match up to the finer grids.

We conclude this discussion
by testing the sensitivity
of the agreement of the number of simulated halos
with the observations for the different P$^3$M simulations
(Fig.~15)
and then by studying the evolution
of the number of the simulated halos (Fig.~16).

In Fig.~15 we show $N(V_{\rm circ})\Delta V_{\rm circ}$
for CDM12($64^3$,51.2,52; $\epsilon=40$ kpc comoving)
(top panel) and
for CDM16($144^3$,100,85; $\epsilon=65$ kpc comoving)
(bottom panel), both analyzed with a $512^3$ DENMAX grid.
We extract several facts from Fig.~15.
First, the trend of increasing number of halos with increasing
force resolution is verified comparing the simulations with
$\epsilon=40$ kpc comoving (top panel) and $\epsilon=65$ kpc comoving
(bottom panel), but the
differences are small.  We found in $\S$ 2 that the CMF was higher for higher
mass resolution simulations and for higher force resolution
simulations independently; but here force resolution must be dominating
because the $\epsilon=40$ kpc comoving simulation has slightly lower mass
resolution than
the $\epsilon=65$ kpc comoving simulation, yet still produces slightly more
halos at
a given $V_{\rm circ}$.
We also found in $\S$ 2 that the differences in the CMF
versus mass resolution were much smaller when we imposed a radius
cut on the masses.  This is equivalent to computing circular velocities.

In Fig.~16 we show $V_{\rm circ}$ for CDM16($144^3$,100,85)
using a fixed
physical radius.
We list the epochs
as redshifts, $z=1/a-1$, because here we are studying
the evolution of the halos for a fixed normalization.  We assume that the
present epoch, $z=0$, is $a_0=\sigma_8=1$.
We keep the physical radius cut constant at 100 kpc
by using a $100a_0/a$ kpc comoving radius cut in equation~(3.2).
The vertical axis
is scaled to a $(51.2~{\rm Mpc})^3$ comoving box, however.

We see in the panels of Fig.~16 that $N(V_{\rm circ})\Delta V_{\rm circ}$
(using a fixed proper radius) evolves strongly with redshift.
At $z=9.9$ halos are still forming.
The major era when galaxies begin to take on
the observed distribution is around $z=3.7$ to $2.2$.
Further evolution indicates that the halos are merging, {\it i.e.}
the curves are decreasing.
At intermediate circular velocities ($V_{\rm circ}=200\kms$)
the number of halos decreases by
a factor of 3.7 from the maximum at $z\sim 3.7$ to $z=0$.
For smaller halos ($V_{\rm circ}=150\kms$) the effect
is higher, a factor of 4.7 from the maximum at $z\sim 3.7$ to $z=0$.
{}From $z=0.4$ to $z=0$ the factor is rougly constant
at $\sim 1.5$ over wide range of $V_{\rm circ}$.
The most massive halos grow at the expense of the smaller ones.  For halos with
total bound masses exceeding $2.3\times 10^{13}{M}_\odot$
in CDM16($144^3$,100,85) ({\it i.e.} 1000 particles),
we find 245 halos at $z=1$,
292 at $z=0.4$, and 285 at $z=0.$  Therefore, the number
of massive halos, unlike the lower mass halos, grows little for $z<1$.

The mergers implied by Fig.~16 are interesting in themselves, and
they are important for Paper II where merging forms
massive systems which have a profound effect on galaxy clustering and velocity
statistics.  Frenk et al. (1988) also found merging in their simulations
with decreasing redshift.  There exists some observational evidence for
merging.  Excess counts of faint galaxies (Tyson 1988) compared with
present galaxy populations
suggest the possibility of merging (Guiderdoni \& Rocca-Volmerange 1990;
Cowie, Songaila, \& Hu 1991).
The merger hypothesis is not without controversy, however, for
other possibilities and complications, highlighted by various
authors, include
1) luminosities may evolve more rapidly for faint galaxies than for
bright galaxies
({\it e.g.} Broadhurst, Ellis, \& Shanks 1988);
2) the geometry of the universe may be different from Einstein-de Sitter
({\it e.g.} Fukugita et al. 1990); or
3) the faint galaxies may represent a separate population
({\it e.g.} Efstathiou et al. 1991).

Cowie, Songaila, \& Hu (1991) argue that the faint galaxy
excess is a factor $\sim 4-5$ from
$z\sim 0.25$ to $z=0$ assuming no luminosity evolution for these modest
redshifts.  Although CDM16 predicts only a factor of $\sim 1.5$ from $z=0.4$
to $z=0$, we cannot accurately address galaxy merging with our dark simulations
for the following reasons:
1) we {\it underestimate} merging
by always associating one galaxy per halo and 2) we {\it overestimate}
merging by always assuming that when halos merge their associated galaxies
merge.
Complications aside, since the reduction is $\sim 1.5$ over
a wide range in $V_{\rm circ}$, we may naively assume
that only $\sim (1-1/1.5)\times 0.7=23\%$ of the spirals
(assuming a $70\%$ spiral fraction) have not experienced a
major merger since $z\sim 0.4$.
This is problematic since Toth \& Ostriker (1992) argue
that high merger rates in
the last 5 Gyr (z=0.37 for $\Omega=1$, $H_0=50~\kms~{\rm Mpc}^{-1}$)
can heat disk galaxies beyond observed levels.
Furthermore, if we were to identify $a_0=\sigma_8=0.5$ as the present
day, Fig.~16 would still apply if the $V_{\rm circ}$ values
were all multiplied by $2^{-1/2}$.  This would move
the $\sigma_8=0.5$ curve ($z=1.0$ in the figure) into
agreement with the Schechter function, as it should from
Fig.~15.  From this we conclude that in the CDM model merging should
continue into the future at a rate as prodigious as the recent past
further violating the Toth \& Ostriker (1992) limits.
For a more detailed examination of merging in CDM models,
see Kauffmann \& White (1993).

\goodbreak
\vskip .2truein
\centerline{\it 4.3. Massive Halos: Computational Issues}

Since the number of halos from
CDM16($144^3$,100,85) agrees with the observed
number of halos in the range
\hbox{$150\kms\lsim V_{\rm circ}\lsim 350 \kms$} we now
focus on the discrepancies outside these ranges.
In this section we explore circular velocities at various
radii and we investigate the sensitivity of the formation
of massive halos to dynamic range and to methods for
identifying the halos.  In $\S~4.4$ we
compare the number of simulated halos with high $\sigma_r$
to the number of observed bright ellipticals, followed
by a discussion of low mass halos in $\S~4.5$.
The purpose of {\it this} section
is to reveal which computational effects, and why, affect
the massive halos.

We present the four most massive halos at $\sigma_8=0.5$ from
CDM12($64^3$,51.2,52) in Table~2
and from CDM1($128^3$,51.2,280) in Table~3.
The halos are labeled
A, B, C, and D.  These two simulations use equivalent
initial conditions.  Corresponding halos are identified.
In the tables we list the circular velocities
in $\kms$ using $R=$ 150 kpc comoving, 200 kpc comoving, and 300 kpc
comoving.  We also list
the bound masses ($R<\infty$) in solar masses.  In Table~2
the results are tabulated
for a $512^3$, $256^3$, $128^3$, and $64^3$ DENMAX grid, all at $\sigma_8=0.5$.
In each column we also list a local rank.  The number $n$ means
the halo is the $n^{\rm th}$ largest halo in the catalog
using the method for halo identification mentioned
in the first column.
Note that the circular velocity
profiles for these massive halos are still rising far
beyond the softening scale.  Here we are interested in their
profiles at large radii.  We use $\sigma_r$ and $V_{\rm circ}$
extrapolated to more
reasonable radii in the next section.

We use Table~2 to study the effect of the choice of $R$
and the DENMAX grid on the massive halos.
The first important feature brought out
is that $V_{\rm circ}$ increases with increasing radii.
These massive halos have extended halos with rising circular
velocities at these scales (cf. Fig. 11).
The next trend we observe is that the circular velocities,
unlike the CMF without a radius cut,
are not very sensitive to the choice of DENMAX grid.
However, the slight differences are explained below.

In Fig.~17
the bound particles from halo B
found in CDM12($64^3$,51.2,52)
are shown using the various DENMAX grids.
We see that the coarser DENMAX grids ($\le 256^3$)
merge the massive halo with an additional small
halo (located at $x\approx 200$ kpc comoving, $y\approx -200$ kpc comoving).
The mass of this ``appendage'' is small
and is far enough away from the
core (about 300 kpc comoving) so that it contributes
little to the circular velocity defined within 300 kpc comoving.  Nevertheless,
it reveals substructure present in the higher force
resolution simulation.

The lower resolution DENMAX grids also lead to the inclusion of
more peripheral (distant) particles.
This is not serious since this does not involve a lot of mass and
only involves particles well beyond 300 kpc comoving from the halo core.
When the DENMAX grid is finer than the interparticle
separation in the periphery, the density gradients are not present
to move the particles into the halo.  This fact partially
explains why the CMF (in $\S$ 2), based on total bound
masses, is more sensitive than $V_{\rm circ}$ to variations
of the DENMAX grid.

We now consider the effects of force resolution.
In Fig.~18 we show the same halo B but from the low force resolution
PM simulation CDM1($128^3$,51.2,280)---we show
every eighth particle for comparison with
the $64^3$ particle P$^3$M simulation.  The force resolution is too
low to produce the ``appendage'' that we see in the P$^3$M simulation---
therefore, there is no significant difference between the $512^3$ grid
DENMAX and the $256^3$ grid DENMAX results.
We conclude that high force resolution reveals more substructure
than low force resolution and that high resolution DENMAX
grids are required to reveal this substructure.

We see in the lower right panel of Fig. 18 a major failing of FOF($l$=0.2).
This is
a particularly pathological example.
Of course we could naturally prune
this halo into separate halos.  It is not practical, however,
to examine visually and prune manually the
thousands of halos produced in each simulation.

The CDM1 PM halos A through D, corresponding to the
halos studied in the CDM12 P$^3$M simulation,
are tabulated in Table~3.  The results are shown for two DENMAX
grids and two FOF linking parameters, all at $\sigma_8=0.5$.
{}From Table 3 we conclude:  1) The $512^3$ DENMAX
results compare well with the $256^3$ DENMAX results.
2) The FOF analyses fail to agree with the DENMAX analyses.
The difference between FOF($l$=0.1)
and FOF($l$=0.2) is not too great since these massive PM halos
do not have a lot of substructure.  The exception
is Halo B in the FOF($l$=0.2) analysis.  This is the
pathological halo shown in the lower right panel of Fig.~18.
Three visually distinct halos are merged together and the
center-of-mass is such that the ``halo'' is highly
non-spherical leading to unreliable circular velocities.

We also find, from Table~3, that the P$^3$M halos are more compact than
the PM halos.
If we compare $V_{\rm circ}$ defined at 150 kpc comoving
in Table~2 for P$^3$M CDM12 with $V_{\rm circ}$ defined at
300 kpc comoving in Table~3 for PM CDM1, we find
comparable values of $V_{\rm circ}$.
By the time we go out to 300 kpc
comoving in the PM simulation we pick up enough particles to give the same
circular velocity as the P$^3$M simulation using 150 kpc comoving.
This is because we choose values of
$R$ to be the radius where most of the
circular velocities are flat.  These radii are directly related to
the force resolution.  However, things do not always work out this nicely
for the massive halos that have rising circular velocity profiles,
as we can see by comparing halo D in Table~2 for P$^3$M CDM12 using 150 kpc
comoving
and halo D in Table~3 for PM CDM1 using 300 kpc comoving.  The differences in
circular velocities are significant enough to shift some of
the massive halos into adjacent
25 $\kms$ bins.

We now summarize some effects arising
from the computational techniques
that influence the number of massive halos in the
$N(V_{\rm circ})\Delta V_{\rm circ}$ histograms.
1) The results are
sensitive to the choice of $R$ used to compute
the circular velocities---this is obvious since the circular
velocity profiles are not flat for the massive halos.
2) We have shown that
higher resolution DENMAX grids reveal more substructure in
some of the massive halos found
in the higher force resolution simulations.
However, from the
images it appears that
no obvious substructure is present in many of the massive halos.
3) Lower resolution DENMAX grids include more peripheral
particles in the halos than higher resolution DENMAX grids.
This arbitrary choice of DENMAX grid does not affect
most computations of circular velocities.  It does, however,
affect the computations of total bound masses; this
explains why the CMF is more sensitive to the limitations
of the current version of DENMAX than is the case
for the circular velocities.
4) FOF($l$=0.2) occasionally links together visually distinct halos.
FOF($l$=0.1) and FOF($l$=0.2) produce similar results for many of the
massive halos but they often fail to match up with DENMAX results
which, visually, appear to do a good job in many cases.
5) The P$^3$M simulations produce halos that are more
compact than the PM simulations.  However,
if a larger value of $R$
is chosen for the PM simulations, then the PM circular
velocities agree with the P$^3$M circular velocities
in most cases.

\vskip .2truein
\centerline{\it 4.4. Massive Halos: Simulations versus Observations}

The distribution of simulated halos with circular velocities
in the range $150\kms\lsim V_{\rm circ}\lsim 350 \kms$ is in
reasonable agreement with observations (Fig. 15).
However, there are too many halos with circular velocities
exceeding $350\kms$.
A simulation with increased force resolution
can reveal more substructure in massive halos
and a continuum-limit DENMAX algorithm would be helpful for
analyzing such simulations.
We take the approach, in this paper, that these massive dark matter
halos represent single, large galaxies.
The possibility that they may
represent clusters is studied in detail in Paper II.

We use fairly complete
catalogs of observed bright ellipticals to estimate
their number density.
It is not accurate enough to estimate the brightest, relatively
few elliptical galaxies simply from a Schechter
luminosity function and a Faber-Jackson relationship.  The
problem is exacerbated by the large amount of scatter
relating $\sigma_1$ to $V_{\rm circ}$ for the simulated massive
halos.
In this subsection we instead use $\sigma_r$ to characterize the
simulated massive halos and we compare them with the number
of observed ellipticals using complete elliptical surveys.
We use these comparisons to constrain the normalization
of the $\Omega=1$ CDM power spectrum using the fact
that as the simulations evolve merging creates more massive halos.

We begin by noting that there are observed galaxy halos
that have large measured circular velocities beyond $\sim 100$ kpc.
The giant elliptical galaxy M87 has been studied by many
workers using the X-ray emitting gas to trace the gravitational
potential well ({\it e.g.} Fabricant \& Gorenstein 1983; White \& Sarazin 1988;
Tsai 1994).
Tsai (1994) modeled the X-ray emission from M87 using a
multi-phase gas assumed to be in hydrostatic equilibrium.
Tsai found the best fit
gas temperature and mass density profile
consistent with both X-ray continuum and line
emission data.
His results are consistent with the velocity dispersions of
Sargent et al. (1978) and Mould et al. (1990) on small scales.
(Note that the mean, radial velocity dispersion, $\sigma_r$,
of stars in M87 from 1 kpc to 4.5 kpc is roughly constant
at only $278\pm 11 \kms$, yet can be as high as $350\kms$
well within 1 kpc--cf. Sargent et al.)
The inferred mass within 300 kpc assuming a Hubble constant
of 50 $\kms~{\rm Mpc}^{-1}$, is approximately $2.5\times 10^{13}{
M}_\odot$ with a corresponding circular velocity of
592 $\kms$.  Curiously, $1.1\times 592\kms/\sqrt{3}=376\kms$
(see equation~(3.8)) which is close to the $350\kms$ measurement
(within 1 kpc) from Sargent et al (1978).  However, since it
is not clear which small-scale star measurements should be related to
large-scale dark matter measurements, we adopt an empirical scaling
law which relates Faber et al. (1989) central velocity measurements
(used as a complete catalog of nearby ellipticals)
to Tsai (1994) circular velocity measurements on large scales.

In Fig.~19 we show circular velocity profiles for halos B and C
(see Tables~2 and 3) from the simulations at $\sigma_8=0.5$
and for M87 (Tsai 1994).  We choose halo C
because it has a circular velocity comparable to M87
at large radii.
(Halo B has a higher circular velocity than M87.)
The profiles from PM CDM1 rise slowly which is expected since
the force softening is 280 kpc comoving.
The profiles from P$^3$M CDM12 rise more quickly than PM CDM1 because
of higher force resolution.
Ignoring the fact that many of the simulated halos are still
rising beyond 150 kpc comoving,
the conjecture that at least some of the very massive simulated halos
are similar to objects like M87 is seen
to be plausible.

We offer a possible explanation why the simulated rotation curves are
still rising beyond the softening scale for P$^3$M CDM12 while
M87 has a very flat rotation curve.
During the dissipational formation of M87,
dark matter can be pulled into the central
region by baryonic infall ({\it e.g.} Blumenthal et al. 1986).
If we examine $V_{\rm circ}(r)$ in figure 3 from Blumenthal et al. (though from
a system with maximum circular velocity $\sim 200\kms$)
we estimate that the ratio of
the distances where $V_{\rm circ}$ turns
over is $\sim 80~{\rm kpc}/10~{\rm kpc}~=8$.
For the P$^3$M CDM12 halo C profile shown in Fig.~19, this effect
could possibly ``pull''
the turnover in the dark matter rotation curve from $\sim 200$ kpc
to $\sim 25$ kpc, consistent with the turnover in the M87 profile
shown in Fig.~19.

We examine the largest halos found in the simulations
and we compare them to one of the most massive and luminous galaxies
known---the central cD galaxy in the cluster A2029 (Dressler 1979;
Uson, Boughn, \& Kuhn 1991).  The mass profile of this
galaxy has been estimated with
a 3-component model by Dressler (1979):
1) a ``normal'' elliptical galaxy, 2) an extended
halo of luminous material out to 100 kpc, and 3) a dark cluster-filling
component.  Dressler estimated the mass within
100 kpc (for $H_0=50\kms~{\rm Mpc}^{-1}$) to be $\sim 3.9\times 10^{13}~{
M}_\odot$ and within 1 Mpc to be $\sim 8.3\times 10^{15}~{M}_\odot$.
The evidence that the material within 100 kpc is part of the central cD
galaxy is strong, but there is some controversy about the
mass out to 1 Mpc.
Uson, Boughn, \& Kuhn  have
argued that the material out to 1 Mpc and beyond is indeed
part of the central cD galaxy.  They measure diffuse light out
to several Mpc.  They found that it has an elliptical profile
with the same axis ratio and
orientation as the central cD galaxy,
and that this is different from the distribution of
the cluster galaxies as a whole.

To compare with the above measurements,
we compute the mass within 100 kpc comoving and 1 Mpc comoving
from the simulated halos.
Using CDM12($64^3$,51.2,52)
we find the halo with the largest mass within 100 kpc comoving
and the halo with the largest mass within 1 Mpc comoving.
The results within 100 kpc comoving (more than twice the Plummer softening)
are $9.0\times 10^{12}{M}_\odot$ at $\sigma_8=0.5$,
$1.7\times 10^{13}{M}_\odot$ at $\sigma_8=0.7$, and
$2.9\times 10^{13}{M}_\odot$ at $\sigma_8=1.0$.
None of these are greater than Dressler's estimate for
the central cD galaxy in A2029, $3.9\times 10^{13}{M}_\odot$.
Within 1 Mpc comoving we find
$1.3\times 10^{14}{M}_\odot$ at $\sigma_8=0.5$,
$1.9\times 10^{14}{M}_\odot$ at $\sigma_8=0.7$, and
$1.9\times 10^{14}{M}_\odot$ at $\sigma_8=1.0$.
Again, these are all smaller than Dressler's estimate,
$8.3\times 10^{15}{M}_\odot$.
Thus, we cannot rule out CDM by arguing that it produces halos with
absolutely too much mass.
We also cannot rule out CDM merely by the fact that our
simulation fails to make
at least one halo as massive as the central cD galaxy
in A2029---we sample only a 51.2 Mpc box
while A2029 is at a distance of 470 Mpc.

We examine other simulations in an attempt to find
halos as massive as the central cD galaxy in A2029.
We examine CDM6($256^3$,51.2,190)
at $R$= 1 Mpc comoving at $\sigma_8=1.0$.
The most massive halo at this radius has a mass
of $3.7\times 10^{14}{M}_\odot$.  Also, we examine
CDM16($144^3$,100,85)
at $\sigma_8=1.0$ with no cut in radius, and the most massive halo
has a mass of $8.9\times 10^{14}$.  This
is a larger box with larger waves in the initial conditions
and a different set of initial random numbers.  The
model still fails to produce a halo as massive
as $\sim 8.3\times 10^{15}{M}_\odot$.  Thus
far the $\Omega=1$ CDM model may be safe.

Although we cannot reject CDM based on the most massive halo in the
simulations, maybe we can reject it based on the large number of slightly
less massive halos that are formed.  Because the most massive galaxies
are ellipticals, we compare the number of simulated halos with large
radial velocity dispersion $\sigma_r$ (second of equations (3.1))
with the number of
ellipticals having large line-of-sight central velocity dispersion.
For the observations we use the samples of nearby bright elliptical galaxies
from Faber et al. (1989) and from the Dressler (1991) supergalactic
plane redshift survey.  We count the number of ellipticals in
these samples with $\log_{10}\sigma_r\ge 2.5$ (20 ellipticals).
We then impose a distance cut of $6000\kms$ (based on corrected distances
from column 12 of table 3 from Faber et al. 1989).  This leaves 14 ellipticals.
The samples are fairly complete.  For the range in apparent magnitudes of
our list of ellipticals, the completeness fraction ranges
from $100\%$ for ${B_T}\lsim 11.6$ down to $20\%$ for ellipticals
in the southern sample with ${B_T}\sim 13$.
If we fold in the completeness fractions (figure 2 from Faber et al. 1989)
the number of 14 ellipticals with $\log_{10}\sigma_r\ge 2.5$ within
a distance of $6000\kms$ might be as high as $\sim 23$.

We also estimate the number of observed ellipticals from
the above samples with $\sigma_r\ge 350\kms$.  This allows us to study very
high values of $\sigma_r$, for which the completeness fractions are much
higher.  Within $6000 \kms$ there are only three known galaxies with
$\sigma_r\ge 350\kms$: SPS 1120 ($\sigma_r=382 \kms$; ${B_T}=12.68$),
NGC 507 ($\sigma_r=366 \kms$; ${B_T}=11.63$)
and NGC 4486 (M87; $\sigma_r=361 \kms$; ${B_T}=9.52$).  The completeness
fraction (based on ${B_T}$)
for SPS 1120 is $\sim 30\%$ and the completeness fraction
for the other two objects is $100\%$.  This tightly
constrains the number of observed ellipticals with
$\sigma_r\ge 350\kms$ within $6000\kms$ to 5.

To demonstrate the inaccuracies at the high mass end
associated with methods presented in $\S~3$,
we compare the above complete estimates with the use of the
Faber-Jackson relationship (equation~(3.7)) using $\sigma_1$ (first
of equations~(3.1)) and
the Schechter function described in $\S~3$.
Again, we weight the Schechter function
by $30\%$; {\it i.e.} we only estimate the elliptical
contributions.
We find in a spherical volume of radius 120 Mpc comoving: 39 objects
with $\sigma_1\ge 316\kms$
and 11 objects
with $\sigma_1\ge 350\kms$.
These numbers are about a factor of 2 larger than the estimates given
above, suggesting that our assumed Faber-Jackson relation
underestimates the luminosity and/or that
ellipticals make up less than $30\%$ of bright galaxies.
(We combine ellipticals and lenticulars in getting the population
fraction of 30\%, but lenticulars are underrepresented among the most
massive galaxies.)
In any case, these results suggests that we
have overestimated the observed $N(V_{\rm circ})$
in Figs.~13 through 16 for $V_{\rm circ}\gsim 500\kms$,
making the disagreement with the simulations even
worse.

To make a better comparison of the simulations with observations, we
estimate the number of simulated halos from four simulations
with $\sigma_r\ge 316\kms$ and $\sigma_r\ge 350\kms$.
The results are shown in Table~4 for simulations with a
variety of force resolution and mass resolution.
All numbers are
scaled to a $(51.2~{\rm Mpc})^3$ comoving volume.
The observations are shown as OBS.~I (without completeness fractions
folded in) and as OBS.~II (with completeness fractions folded in).
All simulations
use a 51.2 Mpc box except for CDM16 which
uses a 100 Mpc box.
The initial conditions
for CDM6, CDM1, and CDM12 are all generated from the same
set of $256^3$ random numbers.  The initial
conditions for CDM2--5 and CDM16
are all generated from different sets of random numbers.
We also show averages, with $1\sigma$ fluctuations,
computed from CDM1--5.

We estimate the velocity dispersions from the simulated halos
in two ways: $\sigma_r$ and $\widetilde\sigma$ which we
describe below.  We compute the radial velocity dispersion,
second of equations~(3.1), within a radius listed in the footnotes of Table~4.
We count the number of halos with $\sigma_r$ exceeding
$316\kms$ and $350\kms$.  We also try using $\sigma_1$ (not shown),
first of equations~(3.1), and the results are
similar to the results using $\sigma_r$ (the differences arise
from the fact that $\sigma_1$ is typically $\sim 20\%$ lower
than $\sigma_r$ as mentioned earlier).

The high velocity dispersions
of the dark matter may not correspond to the velocity
dispersions of the optical galaxies expected to
be embedded well within the cut-off
radii used here.  As a crude estimate of
the central velocity dispersion of a galaxy expected
to be embedded in the simulated halo, we use the following:
$\widetilde\sigma\equiv {361\kms}\times V_{\rm circ}(R)/
600\kms$.
We choose this because the central velocity dispersion
of M87 is $361\kms$ from Faber et al. (1989) and the
estimated circular velocity profile for M87 (from Tsai 1994;
see Fig.~19) is about 600 $\kms$ for $R\gsim 100$ kpc.
Admittedly this is a crude estimate, but it is a simple
attempt to use a single, well-measured object to scale
the simulated data, and it serves as a conservative
check for our comparisons with observations.
(Note, in this case equation~(3.8) works fairly well for $F=1.1$.
However, it may not work well for all objects.)

In Table~4 we compare the results from the simulations
with the observations.
The most important conclusion is that all cases
for $\sigma_8\gsim 0.4$ yield far more halos than
the observed numbers.  The epoch $\sigma_8=0.3$ is not ruled out
since it is difficult to make conclusions
based on zero or one halo.
The results at $\sigma_8=0.4$ indicate that
there are too many halos with $\sigma_r\ge316\kms$.
The problem is less severe (yet not an order of magnitude different from
$\sigma_r$)
using $\widetilde\sigma$,
but the observations tell us that even a single halo
with $\sigma_r\ge350\kms$
in a 51.2 Mpc box is too high by at least a factor of 10.

White et al. (1987),
at $\sigma_8=0.4$ using the same normalization of the CDM power
spectrum as we do, found a single halo
with a circular velocity exceeding 800 $\kms$ in a 50 Mpc
box from three simulations, corresponding to 0.36 halos
for a single 51.2 Mpc simulation.  Our CDM12 simulation has
nearly identical force and mass resolution.
Our largest halo in CDM12 at $\sigma_8=0.4$
has a circular velocity of 567 $\kms$ defined at 100 kpc comoving.
We cannot safely rule out $\sigma_8=0.4$ particularly
since results using $\widetilde\sigma$ only reveal
one very massive halo.
The problem becomes rapidly worse for larger $\sigma_8$;
$\sigma_8\gsim 0.7$ predicts more than 20 times too many
galaxies with $\sigma_r\ge 350\kms$.
This is a severe
problem for $\Omega=1$ CDM since estimates of $\sigma_8$
based on
clustering typically require $\sigma_8\gsim 0.4$ (Davis et al. 1985;
Park 1990; Couchman \& Carlberg 1992).

We now compare the different simulations with each other.
We see the general trend, in Table~4, that both an increase in
mass resolution and an increase in force resolution increase the
production of massive halos.  For the simulations using
equivalent initial conditions (CDM6, CDM12, and CDM1)
we examine corresponding massive halos.
The higher force resolution simulations produce more
compact halos than the lower force resolution simulations;
the cut-off radius is chosen to compensate
for this fact
for reasons discussed earlier.  In some cases, however,
the cut-off radius does not compensate for
the compactness of the high force resolution halos.
We also find that
the high force resolution halos have higher central velocity dispersions.

An increase in mass resolution also increases
the production of massive halos.  The effect is strongest
at $\sigma_8=1.0$ where the numbers from the $256^3$ particle simulation
are far higher than the other simulations compared with earlier
epochs. We observe that the results from
the 100 Mpc box simulation
CDM16 with good mass and force resolution are in reasonable
agreement with the other simulations except at $\sigma_8=1$.
We also observe that the $128^3$ particle, $R_{1/2}=280$ kpc
comoving PM simulations
produce the smallest number of massive halos---these simulations
rank low in the combination of force and mass resolution
and $R=300$ kpc comoving is too close to
$R_{1/2}=280$ kpc comoving.

We conclude that the $\Omega=1$ CDM model is in serious trouble.
The simulations produce far too many massive halos
and an increase in force and mass resolution only make matters worse.
We are able to rule out all normalizations of the primeval density
fluctuations with $\sigma_8\gsim 0.5$.
Using complete catalogs of nearby bright ellipticals, we have constrained the
CDM model more convincingly than by using
the luminosity function at the bright end (cf. Fig.~15 and
Frenk et al. 1988).
The case against $\sigma_8=0.4$ is not as strong as the case
against $\sigma_8\gsim 0.5$.  We found, at the very least, a single
halo with an estimated central velocity dispersion
exceeding $350\kms$ in a single 51.2 Mpc box simulation.
The observations predict that we should only find one such
object in no fewer than 11 simulations.

We know that the simulations suffer from the overmerging of massive halos.
Gas dynamical dissipation could reduce the merging of galaxies.
The result might be to prevent the formation of excessively massive
galaxies, although we consider this unlikely because dissipation should
only increase the central concentration of mass in the most massive halos,
thereby increasing further the central velocity dispersions.  Also, if
the most massive halos actually should represent clusters of galaxies,
then these clusters must still have the correct multiplicity function
(distribution of richness).  Bahcall \& Cen (1992) concluded that the
CDM model with $\sigma_8=1.05$ produces an order of magnitude too many
rich clusters.  In Paper II we investigate the cluster multiplicity
function in detail using our own high-resolution N-body simulations.

\vskip .2truein
\centerline{\it 4.5. Low Mass Halos}

We now examine the low mass halos.  We found
earlier that the P$^3$M simulations produce
too many halos with $V_{\rm circ}\lsim 150 \kms$
(see Fig.~15).  Frenk et al. (1988) argue
that the number of halos is in reasonable agreement
with the observations down to about $60\kms$
using 32000 particle P$^3$M simulations in 14 Mpc boxes.
However, they warned the reader that simulations in larger volumes
predict too many halos (White et al. 1987).
The particle mass in the Frenk et al. simulations
is $5.8\times 10^9{M}_\odot$ and the force resolution
is $\epsilon=14$ kpc.
We have two P$^3$M simulations with lower mass and
force resolution that are computed in a 51.2 Mpc box and a 100 Mpc box
giving us better statistics:
CDM12($64^3$,51.2,52; $m_{\rm part}=3.5\times 10^{10}{
M}_\odot$;
$\epsilon=40$ kpc comoving) and
CDM16($144^3$,100,85; $m_{\rm part}=2.3\times 10^{10}{
M}_\odot$;
$\epsilon=65$ kpc comoving).  Using these simulations,
we explore the effects of resolution and we re-examine
the observational data at the low mass end in order to
explore the apparent excess number of low mass halos compared
with the observations.

The smallest galaxies for which there are reliable mass estimates
have $V_{\rm circ}$ down to about 50 $\kms$  (see Kormendy 1991
and references therein).  Halos from
the 40 kpc comoving Plummer simulation CDM12  with a cut-off radius
of 100 kpc comoving (roughly twice
the Plummer softening) and 5 particles have a circular velocity
of $87\kms$.  Halos from the 65 kpc comoving Plummer simulation
CDM16 with a cut-off radius
of 150 kpc comoving and 5 particles have a circular velocity
of $70\kms$.  Therefore, we can only study halos down
to $70\kms$ using the P$^3$M simulations.

Halos from the PM simulation CDM1($128^3$,51.2,280)
with 5 particles and a cut-off radius of 300 kpc comoving have
a circular velocity of $18\kms$, and halos from
the PM simulation CDM6($256^3$,51.2,190)
with 25 particles and
a cut-off radius of 300 kpc comoving
have a circular velocity of $14\kms$.  However, these PM simulations
have poor force resolution.  We show that higher force
resolution increases the number of low mass halos.  Therefore it is
misleading to compare the number of low mass halos with the
observations using the PM simulations.

Another problem stems from the fact that we need
to use large cut-off radii to characterize the circular velocities
in the PM simulations.
In order for a galaxy to undergo ``complete collapse'' in a spherical,
$\Omega=1$ model,
it has to have an over-density exceeding
$\delta_g=\delta\rho/\rho\sim 170$ (Gunn \& Gott 1972; BBKS).
A simple calculation shows that this places a lower limit
on the circular velocity for a given cut-off radius $R$.
The circular velocity within $R$ for a density
$\rho$ is simply
$$V_{\rm circ}(R)=
{\left({
{G{4\over 3}\pi R^3\rho\over
R}
}\right)}^{1/2}~.\eqno(4.1)$$  If we demand that
the over-density exceed $\delta_g=\rho/\rho_{\rm crit}-1$,
where $\rho_{\rm crit}$ is the density for an $\Omega=1$
universe given by $3H_0^2/(8\pi G)$,
we get the minimum allowed
circular velocity: $$V_{\rm circ}(R)=
\left({{\delta_g+1}\over 2}\right)^{1/2}H_0R~.\eqno(4.2)$$
For $\delta_g=200$ (close to the critical value, chosen to
yield a simple formula),
$H_0=50\kms~{\rm Mpc}^{-1}$,
and a comoving cut-off radius $R$ measured in kpc, we arrive at the simple
formula for the minimum allowed circular velocity in $\kms$:
$$V_{\rm circ}(R)\approx {1\over 2} \left({R\over~{\rm
kpc}}\right)~
\kms~.\eqno(4.3)$$
Eq.~(4.3) puts a severe limit, $V_{\rm circ}\gsim 150\kms$,
on the PM simulations
that require $R=300$ kpc comoving.
For the P$^3$M simulations that require $R=$
100 kpc comoving and 150 kpc comoving, the restrictions
are $50\kms$ and $75\kms$ respectively.

Before exploring the simulations, we need to examine
the observational parameters used for the Tully-Fisher relationship
and the Schechter luminosity function (see $\S$ 3) for
faint galaxies.  Since we have already shown
that the simulations appear to produce too many halos at
the low mass end, we conservatively choose parameters that produce
the largest number of low mass halos allowed within
the uncertainties of the observations.
(We find that there are still too many halos predicted by
the CDM model so we are not forcing the observations
to agree with the model---we are simply estimating how
significant is the discrepancy.)
In the following discussion we re-scale all relevant
numbers to a Hubble constant $H_0=50\kms~{\rm Mpc}^{-1}$.

First we consider the Tully-Fisher relationship in
equation~(3.6).  Pierce \& Tully (1988) reported that
the scatter in this relationship
is $\pm 0.25$ magnitudes.
In their fits (figure 9 from their paper),
they found that the faintest galaxy studied,
$M_{B_T}\approx -16+5\log_{10}(50/85)\approx -17.2$,
is slightly brighter than predicted by their best fit.
Alternatively, if one measures the circular velocity
of this faint galaxy,
the Pierce \& Tully relationship would predict that
the galaxy is fainter than it actually is.  Since
the luminosity function is an increasing function of
decreasing luminosity, one would overestimate the number
of faint galaxies.  We take an extreme point of
view.  We will use equation~(3.6) as is with
an added value of 0.5 magnitudes---this is twice
the reported scatter quoted by Pierce \& Tully and
it results in an increase in the estimate for
the number of observed halos as a function of $V_{\rm circ}$.

Next, we consider the luminosity function.
The estimates of Efstathiou, Ellis, \& Peterson (1988) for
the parameters of the luminosity function are estimated
to hold down to about $M_{B_T}\approx -16+5\log_{10}(50/100)\approx -17.5$.
The luminosity function has been studied by previous workers
down to comparably faint magnitudes (see
Felten 1977 for a review).  This faint limit is
comparable to the faint limit of the Tully-Fisher relationship.
Therefore, we use the parameters of the Schechter
luminosity function given in $\S$ 3 but we
use the reported errors to yield the maximum
number of faint galaxies.  We assume
that $100\%$ of the faint galaxies are spirals.  We
use $\Phi^{*}=(1.56+0.34)\times 10^{-2}h^3~{\rm Mpc}^{-3}$ and
$M^{*}_{B_T}=(-19.68-0.10)-2.5\log_{10}h^{-2}$ with $h=1/2$
and we use $\alpha=-1.07-0.05$.

These changes in the Tully-Fisher relationship and the
Schechter luminosity function increase the estimated
number of faint halos with $50\kms\le V_{\rm circ}< 75\kms$
from 373 to 582 galaxies in a $(51.2~{\rm Mpc})^3$ comoving volume.
When we show the number of observed low mass halos,
we use both
the parameters given in $\S$ 3 and the
extremely stretched parameters given
in this section.

In Figs.~20 and 21 we show the low mass end,
$50\kms \le V_{\rm circ}< 200\kms$, from the simulations.
We still use
$25\kms$ wide bins but we re-bin the data
from 50 $\kms$ to 75 $\kms$, 75 $\kms$ to 100 $\kms$, etc.
The observations using the parameters described in
$\S$ 3 with $F=1$ are shown as solid squares.
The ``maximum'' number of faint halos allowed by
the observations minus the default values is used for
the $\pm$ error bars (the asymmetry is because we use logarithms
on the vertical axes; note these are not $1\sigma$ error bars).

The figures list the various simulation parameters and
the choices of $R$.
The simulations are shown down to circular velocities
such that the bins are complete given the mass resolution
limit.
These restrictions exceed the restrictions based on
the over-density argument,  equation~(4.3), for the P$^3$M simulations.
The PM simulations are restricted by
the over-density argument to $V_{\rm circ}\gsim 150\kms$.
The PM simulations produce fewer low mass halos than the P$^3$M simulations.
This must not be taken to mean better agreement;
instead it is an example of how poor force resolution can give
misleading results.

The results for the $\epsilon=40$ kpc comoving Plummer simulation,
CDM12, and the $\epsilon=65$ kpc comoving Plummer simulation,
CDM16, are in reasonable agreement with each other above 100 $\kms$.
CDM16 has slightly
more power on small scales, $\lambda_{\rm Nyquist}={2\pi/k_{\rm Nyquist}}
=2\times (100~{\rm Mpc}/144)$, than does
CDM12, $\lambda_{\rm Nyquist}=2\times (51.2~{\rm Mpc}/64)$.
We learned from the CMF studies ($\S$~2.3)
that small-scale waves affect the low mass end.

In Fig.~21 we show results from CDM12 for $R=100$ kpc
comoving
and 150 kpc comoving at $\sigma_8=0.3$, 0.4, 0.5, 0.7, and 1.0.
We see that the results are not very sensitive to $R$.
In all cases, there are still too many halos particularly
below 125 $\kms$ and definitely below $100 \kms$.
We see that the number of low mass halos, unlike the high mass
halos, decreases with increasing expansion factor
(both effects are due to merging).

We now compare a few numbers at $\sigma_8=0.4$ and
$\sigma_8=1.0$ from CDM12 (with $R=$
150 kpc comoving) for the ranges
$75\kms \le V_{\rm circ}< 100 \kms$ and $100\kms\le V_{\rm circ}< 125
\kms$.  The numbers in these bins from the simulation
are 1087 and 495 respectively for $\sigma_8=0.4$
and 724 and 333 respectively for $\sigma_8=1.0$.
Using the observational
parameters from $\S$ 3 we find 240 and 168 respectively.
Using the extreme observational parameters discussed in this
section boosts the numbers to 360 and 247 respectively.
Therefore, the excess number of halos below $V_{\rm circ}\sim 125\kms$
is significant.  The simulations produce factors $\sim 2-3$
too many faint halos.
As a final check, we use the parameters described in
this section but we try $\alpha=-1.25$ which boosts
the number for $75\kms\le V_{\rm circ}< 100 \kms$ to 553---
still short of the 724 to 1087 found in the simulation.

We conclude that the $\Omega=1$ CDM model produces too many low mass
halos compared with the observations for $V_{\rm circ}\lsim 125\kms$.
We have compared the numbers from a 40 kpc comoving Plummer simulation
with the largest estimates allowed by the observations and
the discrepancy is still large (about a factor of 2).  Increased force
resolution and increased small-scale power in the initial
conditions make the disagreement worse.
Although these disparities are large, Dekel \& Silk (1986)
argued that supernovae in dwarf galaxies can cause
significant gas loss, and therefore dim the galaxies
with small $V_{\rm circ}$.  Perhaps the Tully-Fisher relation breaks
down at such small $V_{\rm circ}$ (though there is little indication
of this in the data of Pierce \& Tully 1988).
For these reasons, though, we consider the excessive number of low mass halos
in the CDM model to be less serious than the excessive number of
high mass halos.

\vskip .3truein
\centerline{\bf 5. CONCLUSIONS}

A promising result for the CDM model is that the
distribution of halos as a function of circular velocity
agrees rather well with the observations
for circular velocities in the range $150\kms$ to $350\kms$.
The agreement is better over this range for the
P$^3$M simulations versus the lower force resolution
PM simulations and the agreement is not very sensitive
to a Plummer softening of 40 kpc comoving versus 65 kpc
comoving over this range.  However, we
found serious problems outside of this range
and the problems are made worse by increasing the
force resolution and the mass resolution.
Although CDM16
is not the highest resolution simulation, it is computed
in a 100 Mpc box;  we will discover in Paper II that
51.2 Mpc boxes are too small to accurately study
clustering.
On the other hand, the properties of individual halos are
not very sensitive to the differences between
a 51.2 Mpc box and a 100 Mpc box---this fortunate fact
allowed us to use many of the 51.2 Mpc box simulations to
explore
effects arising from varying mass and force resolution and from
different methods for
identifying halos.

We now summarize the chief conclusions found in the preceding sections.

{1.} We studied the cumulative mass fraction CMF($M$), the fraction of
all the mass in halos more massive than $M$.
We found the following: (a) We need to compare
the CMF from simulations analyzed with the same effective DENMAX
resolution---lower resolution grids include more peripheral
particles, increasing the total masses of the halos;
(b) The simulation--to--simulation scatter
is small except for the most massive halos;
(c) Higher mass resolution and higher force resolution
each increase the CMF independently.
The effect of increased mass resolution on the CMF is
reduced if we impose a distance cut from the density peaks
of the halos, and comparisons of simulations
with $64^3$, $128^3$, and $256^3$ particles indicate
that the convergence of the CMF with such a cut is plausible.
(d) Small-scale waves in the initial conditions have a very
small effect on the CMF except for the smallest halos;
(e) Long waves (with wavelength exceeding 51.2 Mpc comoving) in the initial
conditions do not affect the CMF for amplitude $\sigma_8\le0.5$;
and (f) The Press-Schechter theory with $\delta_c=1.68$ predicts too many
massive halos and a more rapid growth of the CMF than found in the
simulations.  Substructure within halos is apparently not erased as
rapidly as implied in the Press-Schechter theory.

{2.} Simulated halos generally have mass distributions characterized by
flat rotation curves extending from about two softening radii to 500
kpc comoving
or more.  The most massive halos have shallower density profiles, resulting
in rising rotation curves.  The independence of circular velocity with
radius for most halos allows us to compare simulated halos at radii of
150--200 kpc comoving (in the P$^3$M simulations) with real spirals at
10 kpc comoving or less.

{3.} The distribution of circular velocities of simulated halos
was compared with observations.  We noted above the good agreement
for $150\kms\lsim V_{\rm circ}\lsim 350\kms$ for any of the three
normalizations $\sigma_8=0.5$, 0.7, and 1.0.
In the analysis, {\it for this range in circular velocities},
we found the following:
(a) The agreement with the observations is best
for the P$^3$M simulations and is not very sensitive
to simulations with a Plummer softening of 40 kpc comoving
versus 65 kpc comoving.
(b) The results from
DENMAX agree better with FOF($l$=0.2) than with FOF($l$=0.1).
(c) The distribution of circular velocities, unlike the CMF, is not very
sensitive to the DENMAX grid, but higher resolution
grids are required to pick out substructure in
the P$^3$M simulations. (d) The number of halos characterized
by their circular velocities (using a fixed, physical radius)
indicates, if $\sigma_8=1$
is the present epoch,
that the galaxy mass function takes on its present shape
by $z\sim 3.7$.  Between this epoch and $z=0$, merging reduces the
number of halos by about a factor of 3.7.  Merging is predicted to
continue into the future.

{4.}
We conclude from the studies of massive halos that
the $\Omega=1$ CDM
model is in trouble if these systems represent individual
galaxy halos.  We are able to rule out
normalizations of the primeval density fluctuations
with $\sigma_8\gsim 0.4$ based on the number of massive halos
if the halos represent individual galaxies, although the lower limit
for $\sigma_8$ is uncertain.
We compared the simulations not only with the observed
luminosity function, but also with complete samples of bright
nearby ellipticals.  These observations constrain the model to
$\sigma_8\lsim 0.5$.
We cannot rule out CDM based on the most massive
halo---we do not find any halos at any epochs with
masses exceeding the inferred mass of the central cD
galaxy in A2029.  If the massive halos represent unresolved
clusters, with the central galaxy having a smaller central velocity
dispersion than the surrounding halo, we may relax these constraints.
We consider this possibility further in Paper II.

{5.} We conclude from the low mass
studies that the $\Omega=1$ CDM model produces too many low mass
halos (by factors $\sim 2-3$)
compared with the observations for $V_{\rm circ}\lsim 125\kms$.
The number of faint halos decreases with increasing $\sigma_8$
because of merging.  Nevertheless, the excess is significant
even at $\sigma_8=1$ using extreme assumptions about
the observational uncertainties.
We do not find reasonable agreement down to $\sim 60\kms$ as reported by
Frenk et al. (1988).  Gas loss in dwarf galaxies (Dekel
\& Silk 1986), however, might dim a significant
number of dwarf galaxies, making this problem less critical for
CDM than the high mass problem.

\vskip .3truein
\centerline{\bf ACKNOWLEDGEMENTS}

This research was conducted using the Cornell National Supercomputer
Facility, a resource of the Center for Theory and Simulation in
Science and Engineering at Cornell University, which receives major
funding from the National Science Foundation and IBM
Corporation, with additional support from New York State
and members of its Corporate Research Institute.  We appreciate
the IBM 3090 programming assistance of CNSF consultant Paul Schwarz.  We thank
John Tsai for providing his estimated mass profile for M87 in advance
of publication.  We thank Paul Schechter for useful suggestions.
This work was supported by NSF grant AST90-01762 and in
part by the DOE and NASA at Fermilab through grant NAGW-2381.
\vfill \eject

{\parindent 0pt
\centerline{\bf REFERENCES:}
{\pp Adams, F.C., Bond, J.R., Freese, K., Frieman, J.A., \&
Olinto, A.V. 1993, Phys. Rev. D, \hbox{\hskip0.5truein} 47, 426}
{\pp Bahcall, N., \& Cen, R. 1992, ApJ, 398, L81}
{\pp Bardeen, J. M., Bond, J. R., Kaiser, N., \& Szalay, A. S. 1986,
ApJ, 300, 15 (BBKS)}
{\pp Bertschinger, E. 1991,
After the First Three Minutes, ed. S. Holt, V.
Trimble, \&\hfil\break
\hbox{\hskip0.5truein} C. Bennett (New York: American
Institute of Physics), 297}
{\pp Bertschinger, E. \& Gelb, J. M. 1991, Computers in Physics, 5, 164}
{\pp Blumenthal, G. R., Faber, S. M.,
Flores, R., \& Primack J. R. 1986, ApJ, 301, 27}
{\pp Bond, J. R., Cole, S., Efstathiou, G. \& Kaiser, N. 1991, ApJ, 379, 440}
{\pp Brainerd, T. G. \& Villumsen, J. V. 1992, ApJ, 394, 409}
{\pp Broadhurst, T. J., Ellis, R. S., \& Shanks, T. 1988, MNRAS, 235, 827}
{\pp Carlberg, R. G. \& Couchman, H. M. P. 1989,
ApJ, 340, 47}
{\pp Carlberg, R. G., Couchman, H. M. P., \& Thomas, P. 1990,
ApJ, 352, L29}
{\pp Cen, R. Y. \& Ostriker, J. P. 1992a, ApJ, 393, 22}
{\pp Cen, R. Y. \& Ostriker, J. P. 1992b, ApJ, 399, L113}
{\pp Couchman, H. M. P. 1991, ApJ, 368, L23}
{\pp Couchman, H. M. P. \& Carlberg, R. G. 1992, ApJ, 389, 453}
{\pp Cowie, L. L., Songaila, A., \& Hu, E. M. 1991, Nature, 354, 460}
{\pp Davis, M., Efstathiou, G., Frenk, C. S., \& White, S. D. M. 1985,
ApJ, 292, 371}
{\pp Dekel, A. \& Silk, J. 1986, ApJ, 303, 39}
{\pp Dressler, A. 1979, ApJ, 231, 659}
{\pp Dressler, A. 1980, ApJ, 236, 351}
{\pp Dressler, A. 1991, ApJS, 75, 241}
{\pp Dubinski, J. \& Carlberg, R. G. 1991, ApJ, 378, 496}
{\pp Efstathiou, G., Bond, J. R., \& White, S. D. M. 1992, MNRAS, 258, 1p}
{\pp Efstathiou, G., Ellis, R. S., \& Peterson, B. A. 1988,
MNRAS, 232, 431}
{\pp Efstathiou, G., Frenk, C. S., White, S. D. M., \&
Davis, M. 1988, MNRAS, 194, 503}
{\pp Efstathiou, G., Bernstein, G., Katz, N., \& Guhathakurta, P. 1991, ApJ,
L47}
{\pp Evrard, A. E., Summers, F. J., \& Davis, M. 1994, ApJ, 422, 11}
{\pp Faber, S. M. \& Jackson, R. E. 1976, ApJ, 204, 668}
{\pp Faber, S. M., Wegner, G., Burstein, D., Davies, R. L.,
Dressler, A., Lynden-Bell, D., \& \hfil\break
\hbox{\hskip0.5truein} Terlevich, R. J. 1989,
ApJS, 69, 763}
{\pp Fabricant, D. \& Gorenstein, P. 1983, ApJ, 267, 535}
{\pp Felten, J. E., 1977, AJ, 82, 861}
{\pp Franx, M., Illingworth, G., \& Heckman, T. 1989, ApJ, 344, 613}
{\pp Frenk, S. F, White, S. D. M., Davis, M., \& Efstathiou, G. 1988,
ApJ, 327, 507}
{\pp Fukugita, M., Takahara, F., Yamashita, K., \& Yoshii, Y. 1990, ApJ, 361,
L1}
{\pp Gelb, J. M. 1992, M.I.T. Ph.D. thesis}
{\pp Gelb, J. M. \& Bertschinger, E. 1994, preprint (Paper II)}
{\pp Guiderdoni, B. \& Rocca-Volmerange, B. 1990, AA, 227, 362}
{\pp Gunn, J. E. \& Gott., J. R. 1972, ApJ, 176, 1}
{\pp Hockney, R. W. \& Eastwood, J. W. 1982, Computer
Simulation Using Particles\hfil\break
\hbox{\hskip0.5truein} (New York: McGraw-Hill)}
{\pp Hoffman, Y. \& Shaham, J. 1985, ApJ, 297, 16}
{\pp Holtzman, J. A. 1989, ApJS, 71, 1}
{\pp Katz, N., Hernquist, L., \& Weinberg, D. H. 1992, ApJ, 399, L109}
{\pp Katz, N. \& White, S. D. M. 1993, ApJ, 412, 455}
{\pp Kauffmann, G. \& White, S. D. M. 1993, MNRAS, 261, 921}
{\pp Kormendy, J. 1990, Evolution of the Universe of Galaxies,
(Astronomical Society of\hfil\break
\hbox{\hskip0.5truein} the Pacific: California), 33}
{\pp Kundi\'{c}, T. 1991, M.I.T. S.B. thesis}
{\pp Little, B., Weinberg, D. H., \& Park, C. 1991, MNRAS, 253, 295}
{\pp Melott, A. L. 1990, Phys. Rep., 193, 1}
{\pp Mould, J. R., Oke, J. B., De Zeeuw, P. T. \& Nemec, J. M. 1990, AJ, 99,
1823}
{\pp Park, C. 1990, MNRAS, 242, 59}
{\pp Pierce, M. J. \& Tully, B. 1988, ApJ, 330, 579}
{\pp Postman, M. \& Geller, M. J. 1984, ApJ, 281, 95}
{\pp Press, W. H. \& Schechter, P. 1974, ApJ, 187, 425}
{\pp Rubin, V. C., Burstein, D., Ford, W. K., \& Thonnard, N. 1985, ApJ,
289, 81}
{\pp Sargent, W. L. W., Young, P. J., Boksenberg, A., Shortridge, K.,
Lynds, C. R., \&\hfil\break
\hbox{\hskip0.5truein} Hartwick, F. D. A. 1978, ApJ, 221, 731}
{\pp Schechter, P. L. 1976, ApJ, 203, 297}
{\pp Toth, G. \& Ostriker, J. P. 1992, ApJ, 389, 5}
{\pp Tsai, J. C. 1994, ApJ, 423, 143}
{\pp Tully, R. B. \& Fisher, J. R. 1977, AA, 54, 661}
{\pp Tully, R. B. \& Fouque, P. 1985, ApJS, 58, 67}
{\pp Tyson, J. A. 1988, AJ, 96, 1}
{\pp Uson, J. M., Boughn, S. P., \& Kuhn J. R., ApJ, 369, 46}
{\pp Warren, M. S., Zurek, W. H., Quinn, P. J., \& Salmon, J. K. 1991,
in After the First\hfil\break
\hbox{\hskip0.5truein} Three Minutes, ed. S. Holt, V.
Trimble, \& C.  Bennett (New York: American\hfil\break
\hbox{\hskip0.5truein} Institute of Physics), 216}
{\pp White, S. D. M., Davis, M., Efstathiou, G., \& Frenk, C. S.
1987, Nature, 330, 451}
{\pp White, S. D. M. \& Rees, M. 1978, MNRAS, 183, 341}
{\pp White, S. D. M. \& Sarazin, C. L. 1988, ApJ, 335, 688}
{\pp Wright, E. L. et al. 1992, ApJ, 396, L13}
{\pp Zel'dovich, Ya. B. 1970, AA, 5, 84}
}
\vfill
\eject
\parindent 0pt
\centerline{\bf FIGURE CAPTIONS:}

{\bf FIG.~1:}
Cumulative mass fractions
for CDM1($128^3$,51.2,280)
analyzed using FOF($l$=0.1), FOF($l$=0.2), and $512^3$
DENMAX.
a) Compares
DENMAX (solid curves)
with FOF (dot-dashed curves for $l$=0.1; dashed curves for $l$=0.2)
and b) the effect of the removal
of unbound particles (solid curves for bound particles; dot-dashed curves for
all
particles).
Each case has three curves---lower curves
($\sigma_8=0.5$), middle curves ($\sigma_8=0.7$), and upper curves
($\sigma_8=1$).

{\bf FIG.~2:}
Cumulative mass fractions for $512^3$ DENMAX halos from CDM1--5.
All use $128^3$ particles, a 51.2 Mpc box, and
a force softening distance $R_{1/2}=280$ kpc comoving.

{\bf FIG.~3:}
Cumulative mass fractions for $512^3$ DENMAX halos from various
simulations in 51.2 Mpc boxes.
The effects of particle number, $N$, and force
softening, $R_{\rm 1/2}$, are shown.  All four simulations are generated
from an equivalent set of random numbers.

{\bf FIG.~4:}
Cumulative mass fractions at $\sigma_8=0.5$ for DENMAX
halos from CDM12($64^3$,51.2,52) using
a $512^3$ grid (solid curve), and lower
resolution DENMAX grids:
$256^3$ grid (short-dashed curve); $128^3$ grid (long-dashed curve);
and $64^3$ grid (dot-dashed curve).

{\bf FIG.~5:}
Cumulative mass fractions for $512^3$ DENMAX halos from
two simulations in bigger boxes:  CDM11(128$^3$,102.4,560;
solid curves)
and CDM16($144^3$,100,85; short-dashed curves).
Also, top panel ($\sigma_8=0.5$) for $256^3$ DENMAX halos
from CDM12($64^3$,51.2,52; long-dashed curve).

{\bf FIG.~6:}
Cumulative mass fractions for $512^3$ DENMAX halos
from three $R_{1/2}=280$ kpc comoving
PM simulations in 51.2 Mpc boxes.
All three simulations use equivalent initial conditions.
The dashed curves
are for $128^3$ particles---{\it but} the initial displacements were
interpolated from the $64^3$ particle case (dot-dashed curves).

{\bf FIG.~7:}
Cumulative mass fractions for $512^3$ DENMAX halos at $\sigma_8=0.5$
from three PM simulations using
equivalent initial conditions.
We include only particles within 300 kpc comoving from the density peak.
a) Compares three simulations indicated by particle number
$N$ and force softening $R_{1/2}$.  The halos whose
raw mass (no cut in radius and no unbinding) exceeding
the transition mass
($1.1\times 10^{13}{M}_\odot$; vertical bar)
have not had their unbound particles removed.
b) Results with and without the removal
of unbound particles above the transition mass.
c) The same curves from Fig.~7a except
the $256^3$ particle, $R_{1/2}=190$ kpc comoving simulation has been
scaled to $R_{1/2}=280$ kpc comoving.

{\bf FIG.~8:}
Cumulative mass fractions from CDM12($64^3$,51.2,52) and
CDM16($144^3$,100,85) and
the Press-Schechter theory (PS), all for $\sigma_8=0.5$, 0.7, and 1
with the CMF being larger with increasing $\sigma_8$.
a) Dotted curves are PS with a top hat window function and
$\delta_c=2$.  Solid circles are
CMFs of raw masses for CDM16 and FOF($l=0.2$) while
crosses are CMFs of raw masses for
CDM12 and FOF($l=0.2$).
b) Same as a) except PS is for a gaussian window function with
$\delta_c=2$ and the simulations are CMFs of raw masses
analyzed with FOF($l=0.1$).
c) PS is for a gaussian window function with $\delta_c=2$.
Solid circles are CMFs from CDM12 computed with raw masses
using a $512^3$ grid DENMAX analysis.  Crosses are
CMFs also from CDM12 and also computed using
a $512^3$ grid DENMAX analysis---however, only bound particles
within 200 kpc comoving of the DENMAX peak are used to
compute the CMFs.

{\bf FIG.~9:}
$\sigma_1$ and $V_{\rm circ}\equiv (GM/R)^{1/2}$
from CDM16($144^3$,100,85) at $\sigma_8=0.7$.
Each point represents one halo. We show various comoving cuts.
a) $\sigma_1(200~{\rm kpc})$ versus $\sigma_1(100~{\rm kpc})$.
b) $V_{\rm circ}(200~{\rm kpc})$ versus $V_{\rm circ}(100~{\rm kpc})$.
c) $V_{\rm circ}(300~{\rm kpc})$ versus $V_{\rm circ}(200~{\rm kpc})$.

{\bf FIG.~10:}
$V_{\rm circ}/\sigma_1$ versus $V_{\rm circ}$
(all computed using $R=200$ kpc comoving)
from CDM16($144^3$,100,85).
Solid lines are for $V_{\rm circ}/\sigma_1=\sqrt{3}$,
or $F=1$ in equation~(3.8), and
dashed lines are for $V_{\rm circ}/\sigma_1=\sqrt{3}/1.1$,
or $F=1.1$ in equation~(3.8).

{\bf FIG.~11:}
Circular velocity profiles for $512^3$ DENMAX halos from
CDM12($64^3$,51.2,52; $\epsilon=40$ kpc comoving).
For each halo we computed
$V_{\rm circ}$ at 150 kpc comoving: $V_{150}$.
We then sorted the halos from large to small $V_{150}$.
We show the top ten halos and then every twentieth halo
thereafter, all the way down to $V_{150}=150\kms$.
(This procedure is done independently at each epoch.)

{\bf FIG.~12:}
Cumulative velocity dispersion
profiles $\sigma_1(R)$ for
the same halos shown in Fig.~11.

{\bf FIG.~13:}
Distribution function of circular velocity for
CDM1($128^3$,51.2,280) with
the circular velocities measured at 300 kpc comoving.
The results
are scaled to
a comoving volume
of $(51.2~{\rm Mpc})^3$ in all of the distribution
plots (Figs. 13 through 16) for comparison.
The dot-dashed curves in these plots
are for a Schechter function
($F=1$ and $F=1.1$; the latter gives slightly fewer numbers for
bright elliptical halos).
We use $512^3$ DENMAX (solid histograms),
FOF ($l$=0.1, short-dashed histograms),
and FOF ($l$=0.2, long-dashed histograms).
The histograms, high to low values at $V_{\rm circ}\sim 200\kms$,
are for DENMAX, then FOF ($l$=0.2), then
FOF ($l$=0.1).
For $\sigma_8=1$, each method found two halos for $V_{\rm circ}\ge 800\kms$.
The bins at $V_{\rm circ}=875\kms$ and $1050\kms$
each contain one FOF ($l$=0.1) halo
and one FOF ($l$=0.2) halo.  The bins at $V_{\rm circ}=925\kms$
and $1075\kms$ each contain one DENMAX halo.

{\bf FIG.~14:}
Distribution function of circular velocity
for CDM12($64^3$,51.2,52) analyzed at $\sigma_8=0.5$ with
DENMAX grids of $512^3$ (solid histograms),
$256^3$ (short-dashed histograms),
$128^3$ (long-dashed histograms), and $64^3$ (dot-dashed histograms).
The comoving radii used
to define the circular velocities are: a) 150 kpc, b) 200 kpc,
and c) 300 kpc.
The $512^3$, $256^3$, and $128^3$ DENMAX grids are nearly indistinguishable
except for large $V_{\rm circ}$.  The coarse $64^3$ DENMAX grid fails
to match up to the other histograms.
In the bottom panel, each grid identified two halos
above $700\kms$.  The $800\kms$ bin contains a $64^3$ grid halo
and a $512^3$ grid halo.
The $825\kms$ bin contains a $128^3$ grid halo
and a $256^3$ grid halo.
The $875\kms$ bin contains a $512^3$ grid halo.
The $900\kms$ bin contains a $64^3$ grid halo,
a $128^3$ grid halo, and a $256^3$ grid halo.

{\bf FIG.~15:}
Distribution function of circular velocity
for $512^3$ DENMAX halos from a) CDM12($64^3$,51.2,52)
and from b) CDM16($144^3$,100,85).
We use a 150 kpc comoving distance to compute $V_{\rm circ}$
for CDM12 and a 200 kpc comoving distance to compute
$V_{\rm circ}$ for CDM16.  The results are shown at
$\sigma_8=0.5$ (solid histograms), 0.7 (short-dashed histograms),
and 1.0 (long-dashed histograms); they overlap except
for high $V_{\rm circ}$ where there are more halos
for larger values of $\sigma_8$.

{\bf FIG.~16:}
Same as Fig. 15b, except that we use a fixed physical radius of 100 kpc
to compute circular velocities of halos
in a $(51.2{\rm Mpc})^3$ comoving volume.
The epochs are listed in terms of the redshift $z=1/a-1$ where
we take $a_0=\sigma_8=1$ to be the present day. This plot
depicts evolution.

{\bf FIG.~17:}
The bound particles of
Halo B (see Table~2) from CDM12($64^3$,51.2,52)
at $\sigma_8=0.5$ are shown as found by the various DENMAX
analyses.  The images are shown as $x-y$ projections in units
of comoving kpc.  The $512^3$ DENMAX analysis is able to resolve the small
halo present in the other panels (located at $x\approx 200$ kpc
and $y\approx -200$ kpc).

{\bf FIG.~18:}
The bound particles of
Halo B (see Table~3) from CDM1($128^3$,51.2,280)
at $\sigma_8=0.5$ found by the $256^3$ and $512^3$ DENMAX
analyses and by the FOF($l$=0.1) and FOF($l$=0.2) analyses.
(For FOF we do not remove the unbound particles.)
We show every eighth particle to facilitate a comparison with
the $64^3$ particle simulation shown in Fig.~17.
There is not much difference in the two DENMAX analyses (apart
from the peripheral particles) because the PM forces are computed
on a $256^3$ grid.
The FOF($l$=0.2) analysis reveals a dramatic
shortcoming of FOF---namely the linking together of several
dynamically distinct
halos.

{\bf FIG.~19:}
The circular velocity profiles at $\sigma_8=0.5$
for halos B and C (halo B has a larger $V_{\rm circ}$ than halo C)
from CDM1(solid curves: $128^3$,51.2,280; see Table~3)
and from CDM12(dotted curves: $64^3$,51.2,52; see Table~2).
We show the profile for M87 as computed by
Tsai (1994) based on analysis of X-ray emission (short-dashed curve).

{\bf FIG.~20:}
Distributions of simulated low mass halos from four simulations with various
comoving cuts $R$:
CDM1($128^3$,51.2,280, $R=300$ kpc;
 dot-long-dashed histograms),
CDM6($256^3$,51.2,190, $R=300$ kpc; short-dashed histograms),
CDM16($144^3$,100,85, $R=150$ kpc; long-dashed histograms), and
CDM12($64^3$,51.2,52, $R=100$ kpc; dot-short-dashed histograms).
We estimate the
observed numbers using parameters presented in $\S$ 3 (solid squares)
with plus/minus error bars (these are ``extreme'' systematic errors,
not $1\sigma$ error bars).

{\bf FIG.~21:}
Distributions of simulated low mass halos from CDM12($64^3$,51.2,52).
(The observed numbers are solid squares with error bars, see Fig. 20.)
The results are shown for a) $R=$100 kpc comoving and b)
$R=$150 kpc
comoving.  The results are shown at $\sigma_8=0.3$ (dotted histograms),
0.4 (short-dashed histograms), 0.5 (long-dashed histograms),
0.7 (dot-short-dashed histograms), and 1.0 (dot-long-dashed histograms).
\bye